\documentstyle[apj,times,psfig]{article}

\def\simg{\mathrel{\rlap{\raise 0.511ex \hbox{$>$}}{\lower 0.511ex \hbox{$\sim$}}}}
\def\siml{\mathrel{\rlap{\raise 0.511ex \hbox{$<$}}{\lower 0.511ex \hbox{$\sim$}}}}
\def\veps{\varepsilon} \def\eps{\epsilon} \def\nnbb{nb}
\def\tnu{\tilde{\nu}} \def\tGc{\tilde{\Gamma}_c} \def\ttau{\tilde{\tau}}
\def\ergcm2{\, \rm erg \, cm^{-2}} 
\def\z{{\cal Z}} \def\phi5{\Phi_{-5}} \def\tpm{{\tau}} \def\o{{\rm o}} 
\def\taugg{\tau_{\gamma\gamma}}

\begin{document}

\parskip 5pt
\topmargin 1cm

\title{Synchrotron and inverse-Compton emissions from pairs formed in GRB afterglows
       (analytical treatment)}

\author{A. Panaitescu}
\affil{ Space \& Remote Sensing, MS B244, Los Alamos National Laboratory, Los Alamos, NM 87545, USA}

\begin{abstract}
 We calculate the synchrotron and inverse-Compton emissions from pairs formed in GRB afterglows from
high-energy photons (above 100 MeV), assuming a power-law photon spectrum $C_\nu \propto \nu^{-2}$ 
and considering only the pairs generated from primary high-energy photons. The essential properties 
of these pairs (number, minimal energy, cooling energy, distribution with energy) and of their emission 
(peak flux, spectral breaks, spectral slope) are set by the observables GeV fluence $\Phi (t) = Ft$ 
and spectrum, and by the Lorentz factor $\Gamma$ and magnetic field $B$ of the source of high-energy 
photons, at observer-time $t$. Optical and X-ray pseudo--light-curves $F_\nu (\Gamma)$ are calculated 
for given $B$; proper synchrotron self-Compton light-curves are calculated by setting the dynamics 
$\Gamma(t)$ of the high-energy photons source to be that of a decelerating, relativistic shock. 
It is found that the emission from pairs can accommodate the flux and decays of the optical flashes 
measured during the prompt (GRB) phase and of the faster-decaying X-ray plateaus observed during 
the delayed (afterglow) phase. The brightest pair optical emission is obtained for $100 < \Gamma < 500$,
and depends mostly on the GeV fluence, being independent of the source redshift. Emission from pairs 
formed during the GRB phase offers an alternate explanation to reverse-shock optical flashes. These
two models may be distinguished based on their corresponding flux decay index--spectral slope relations,
different correlations with the LAT fluence, or through modeling of the afterglow multiwavelength data.
\end{abstract}


\vspace*{2mm}
\section{Introduction}

 The first Fermi-LAT GRB catalog (Ackermann et al 2013) identifies a "temporally extended" 
emission at 100 MeV--10 GeV for eight bursts, with four other having LAT detections well after 
the end of the Fermi-GBM prompt phase. The LAT emission of those eight afterglows peaks at 
10--20 s after trigger, having a fluence $\Phi = 10^{-5\pm 1} \ergcm2$, followed by a 
flux decay $\nu F_\nu \propto t^{-1.3\pm 0.3}$ until up to 1 ks, with a photon spectrum
$C_\nu \propto \nu^{-2.1\pm 0.2}$. The energetic output of those LAT afterglows, $E_{100 MeV} =
10^{53 \pm 1}$ erg, is 10--100 percent of their GRB output (at $\sim 1$ MeV). 

 The condition of optical thinness to pair-formation for the highest-energy LAT photons yields 
lower limits on the Lorentz factor $\Gamma$ of their source. Assuming that the prompt LAT emission 
has same origin as the GBM burst and using the burst variability timescale to determine the source 
radius, it was inferred that $\Gamma_{grb} > 200-1000$ (e.g. Abdo et al 2009). The general lack 
of pair-formation signatures in the light-curves and spectra of LAT afterglows indicates that 
the afterglow source also has $\Gamma_{ag} > 200$ (Panaitescu, Vestrand \& Wozniak 2014), with 
a detection bias against afterglows with $\Gamma \siml 75$, for which pair-formation attenuates 
the intrinsic afterglow emission too much and the emergent afterglow is too dim to be detected by LAT.

 In this work, we calculate the synchrotron and inverse-Compton emission from pairs formed in 
LAT afterglows, in a simplified set-up. The temporal and spectral properties of LAT afterglows 
being consistent with those at lower photon energies (optical and X-ray), indicates that the LAT 
emission arises in the forward-shock driven by the GRB ejecta into the circumburst medium 
(Kumar \& Barniol Duran 2009). An important simplification is that we consider only the pairs 
formed {\sl behind} the forward-shock, whose energy distribution is set primarily by the
spectrum of the high-energy photons, and ignore the pairs formed {\sl ahead} of the forward-shock, 
whose energy is set by their shock-acceleration (given that half of the emitted photons travel 
ahead of the forward-shock, it follows that a comparable number of pairs form behind and ahead 
of the shock). 
 For ease of calculating the number of pairs, we assume that the single power-law spectrum of the 
high-energy LAT photons extends well below 100 MeV and above 100 GeV, and that it has a spectral 
index -2 (in photon number). Furthermore, we consider only the pairs formed from high-energy
photons for which the photon front is optically thick to pair-formation. Another simplification 
made is that the threshold energy for pair-formation is set only by the relativistic collimation of 
the seed LAT photons, i.e. we ignore the scattering/decollimation of the high-energy photons by the 
already formed pairs.

 Section \S\ref{npairs} calculates the number of pairs with the above approximations, and their
minimal energy in the shock frame; \S\ref{syic} presents the calculation of the spectral
breaks of the pair emission, the regions in the $nb-\Gamma$ corresponding to various orderings
of the spectral breaks being identified in \S\ref{regions}. 
The calculation of the received synchrotron self-Compton emission, taking into account synchrotron 
self-absorption, radiative cooling, first inverse-Compton scattering if the optical thickness to 
scattering by pairs is $\tpm < 1$, and higher orders inverse-Compton if $\tpm > 1$, is presented
in \S\ref{moresyic}. Optical and X-ray pair light-curves are discussed in \S\ref{results}.

\vspace*{2mm}
\section{Number of pairs and their distribution with energy}
\label{npairs}

 A photon of {\sl observer}-frame energy $\veps$ MeV forms a pair when interacting with another 
photon of energy above the {\sl source}-frame threshold
\begin{equation}
 \eps_t (\veps) = \frac{ 4\Gamma^2 (m_e c^2)^2 }{(z+1) \veps} = 
   35\, \z^{-1} \frac{\Gamma_2^2}{\veps_8} \; {\rm MeV} 
\label{ethr}
\end{equation}
with $\Gamma$ the Lorentz factor of the source that produced both photons, and using the notations
\begin{equation}
 X_n = \frac{X (cgs)}{10^n} \;, \quad \z \equiv \frac{z+1}{3}
\end{equation}
For an afterglow fluence $\Phi$ at 0.1--10 GeV, with a photon-number distribution with energy
$C_\veps \propto \veps^{-2}$, the number of photons with energy above $\eps$ is 
\begin{equation}
 N_\gamma (> \eps) = \frac{4 \pi d_l^2}{z+1} \frac{\Phi}{4.6\, \eps} =
      1.14 \times 10^{56} \z^3 \frac{\phi5}{\eps_8} 
\label{NN}
\end{equation} 
where $d_l = 5.10^{27} (z+1)^2$ cm is the luminosity distance for redshift $z$. 
Thus, the number of photons above the threshold energy $\eps_t$ is
\begin{equation}
 N_\gamma [> \eps_t(\veps)] = 3.3 \times 10^{56} \z^4 \frac{\phi5}{\Gamma_2^2} \veps_8
\end{equation}

 A relativistic source moving at constant $\Gamma$ has radius 
\begin{equation}
 R = \frac{4}{3} \frac{ct}{z+1} \Gamma^2
\label{R}
\end{equation}
at observer-frame time $t$, the factor 4/3 corresponding to photons emitted from the "edge" 
of the source, i.e. from the fluid flowing at an angle $\Gamma^{-1}$ relative to the direction
toward the observer. Thus, the optical thickness to pair-formation for a $\veps$ photon is
\begin{equation}
 \taugg (\veps) = \bar{\sigma}_{\gamma\gamma} \frac{N_\gamma [> \eps_t(\veps)]}{4 \pi R^2} 
      = 1.8\, \z^6 \frac{\phi5}{\Gamma_2^6 t_1^2}  \veps_8
\end{equation}
where $\bar{\sigma}_{\gamma\gamma} = 0.18 \sigma_e$ is the pair-formation cross-section
averaged over the $\veps^{-2}$ photon distribution and $\sigma_e$ is the lepton scattering 
cross-section.
Thus, the afterglow photon front is optically thick to pair-formation ($\taugg >1$)
for photons of energy above
\begin{equation}
 \veps_\pm = 56\, \z^{-6} \frac{\Gamma_2^6 t_1^2}{\phi5} \; {\rm MeV} 
\label{epm2}
\end{equation}

 We approximate the number of pairs formed as that of photons with energy $\veps > \veps_\pm$.
A fraction $\taugg (\veps) < 1$ of the photons with $\veps < \veps_\pm$ also form pairs;
these pairs are roughly a factor $\ln (\veps_\pm/\veps_m) > 1$ more numerous than those for 
which $\taugg (\veps) > 1$, $\veps_\pm$ being the peak energy of the $\nu F_\nu$ LAT spectrum.
This approximation is made for two reasons. One is to avoid carrying unknown observables -- 
$\veps_m$ and the spectral slope below that peak -- in the following calculations, the other
is to work with a single power-law pair distribution with energy. 
The ensuing underestimation of the true pair number increases with $\Gamma$ because $\veps_\pm 
\propto \Gamma^6$. However, for high Lorentz factors, pairs are cooling much faster than they 
are created, and only the pairs formed from photons with $\veps > \veps_\pm$ radiate synchrotron 
emission at the frequencies of interest (optical and X-rays).

 Given that each photon with energy above $\veps_\pm$ yields two leptons (an electron and a positron),
it follows that the number of leptons formed is 
\begin{equation}
 N = 2 N_\gamma(> \eps_\pm) = 1.34 \times 10^{56} \z^8 \frac{\phi5^2}{\Gamma_2^6 t_1^2} 
\label{N2}
\end{equation}
The above results hold when the pair-formation threshold energy for $\veps_\pm$ photons
\begin{equation}
 \veps_t(\veps_\pm) = 21\, \z^4 \frac{\phi5}{\Gamma_2^4 t_1^2} \; {\rm MeV}
\label{et}
\end{equation}
is below $\veps_\pm$, i.e. when there are enough absorbing photons above $\veps_t(\veps_\pm)$.
The condition $\veps_t(\veps_\pm) = \veps_\pm$ defines a Lorentz factor
\begin{equation}
 \Gamma_c \equiv 91\, \z \frac{\phi5^{0.2}}{t_1^{0.4}}
\label{Gc}
\end{equation}
such that $\veps_t(\veps_\pm) < \veps_\pm$ for $\Gamma > \Gamma_c$.

 For $\Gamma < \Gamma_c$, we have $\veps_t(\veps_\pm) > \veps_\pm$, i.e. there there are fewer 
photons above $\veps_t(\veps_\pm)$ than above $\veps_\pm$ and, consequently, not all photons 
above $\veps_\pm$ can form pairs, even though $\tau_{\gamma \gamma}(\veps_\pm) > 1$. 
In this case, the energy $\veps_\pm$ above which all photons form pairs is given by 
$\veps_t(\veps_\pm) = \veps_\pm$. Then, equation (\ref{ethr}) leads to
\begin{equation}
 \veps_\pm = \veps_t(\veps_\pm) = \frac{2 \Gamma m_e c^2}{z+1} = 34\, \z^{-1} \Gamma_2 \; {\rm MeV} 
   \quad (\Gamma < \Gamma_c)
\label{epm1}
\end{equation}
Because the $\veps_\pm$ form pairs mostly with other $\veps_\pm$ photons, the number of 
leptons formed is just the number of photon with energy above $\veps_\pm$:
\begin{equation}
 N = N_\gamma(> \eps_\pm) = 1.11 \times 10^{56} \z^3 \frac{\phi5}{\Gamma_2} \quad (\Gamma < \Gamma_c)
\label{N1}
\end{equation}
using equation (\ref{NN}).


 Most pairs form at threshold, with a typical {\sl lab}-frame energy $\gamma m_e c^2 = (z+1) 
[\veps + \veps_t(\veps)]/2$, thus
\begin{equation}
 \gamma m_e c^2 \simeq (z+1) \left\{ \begin{array}{ll} 
    \veps    &  [\Gamma < \Gamma_c, \veps = \veps_t(\veps)]  \\
    \veps/2  &  [\Gamma \gg \Gamma_c,  \veps \gg \veps_t(\veps)] \\
  \end{array} \right.
\label{lab}
\end{equation}
This approximate one-to-one correspondence between the absorbed photon energy and the 
electron-positron pair energy implies that the distribution with energy of pairs 
is that of the high-energy photons
\begin{equation}
 \frac{dN}{d\gamma} (>\gamma_i) \propto \gamma^{-2}
\label{plaw}
\end{equation}

 The above pair distribution with energy holds above a {\sl shock}-frame energy $\gamma'_i$ 
that can be determined from the pair lab-frame energy $\gamma_i$ corresponding to the minimum 
energy $(z+1)\veps_\pm$ above which all photons form pairs, and from the angle $\theta_\pm$
at which the pairs move (in the lab-frame) relative to the radial direction of shock's motion. 
Taking into account that the high-energy photons are collimated (by the relativistic motion of 
their source, the shock) within an angle $\theta_\gamma \simeq \Gamma^{-1}$ of the radial direction, 
it follows that 
$i)$ the center-of-momentum of the colliding photons moves at angle $\theta_{cm} \sim \Gamma^{-1}$
 relative to the radial direction and
$ii)$ the pairs emerge at a typical angle $\theta_{out} = (\veps \veps_t)^{1/2}/ 
  [\Gamma(\veps + \veps_t)]$ relative to the direction of motion of the center-of-momentum. 
In the lab-frame, the emerging pair moves at an angle $\theta_\pm \simeq \max (\theta_{cm}, 
\theta_{out})$ relative to the shock's direction of motion.

 For $\Gamma < \Gamma_c$, we have $\veps_\pm = \veps_t(\veps_\pm)$, thus $\theta_{out} = 
1/(2\Gamma)$, from where $\theta_\pm \simeq \theta_{cm} \simeq \Gamma^{-1}$, which implies that, 
in the shock-frame (moving outward at Lorentz factor $\Gamma$), the minimal pair energy is 
$\gamma'_i \simeq \gamma_i/\Gamma$. From equations (\ref{epm1}) and (\ref{lab}), 
$\gamma_i m_e c^2 = (z+1)\veps_\pm = 2 \Gamma m_e c^2$, thus 
\begin{equation}
 \gamma'_i = 2 \quad (\Gamma < \Gamma_c)
\label{gi1}
\end{equation}

 For $\Gamma \gg \Gamma_c$: $\veps_\pm \gg \veps_t(\veps_\pm)$, thus $\theta_{out} \simeq 
(\veps_t/\veps_\pm)^{1/2}/\Gamma \ll \Gamma^{-1}$, from where $\theta_\pm \simeq \theta_{cm} 
\simeq \Gamma^{-1}$, which implies that $\gamma'_i \simeq \gamma_i/\Gamma$, as for $\Gamma < \Gamma_c$. 
From equation (\ref{lab}), $\gamma_i m_e c^2 = (z+1)\veps_\pm/2$, which, together with equation
(\ref{epm2}), leads to
\begin{equation}
 \gamma'_i = \frac{(z+1)\veps_\pm}{2\Gamma m_e c^2} = 1.6\, \z^{-5} \frac{\Gamma_2^5 t_1^2}{\phi5} 
     = \left( \frac{\Gamma}{\Gamma_c} \right)^5 \quad (\Gamma \gg \Gamma_c) 
\label{gi2}
\end{equation}


\vspace*{2mm}
\section{Synchrotron and inverse-Compton spectral characteristics}
\label{syic}

 For a power-law distribution with energy of the radiating particles, the synchrotron spectrum 
is a sequence of power-laws with breaks at frequencies
\begin{equation}
 \nu_{i,a,c} = \frac{e}{2 m_e c} \frac{\gamma_i^2 B \Gamma}{z+1} = 
         3.0 \times 10^8 \z^{-1} \gamma_{i,a,c}^2 B_0 \Gamma_2 \; {\rm Hz}
\label{nui}
\end{equation}
where $\gamma$ is the pair random Lorentz factor in the shock's frame (prime notation dropped).
$\nu_i$ is the {\sl injection} frequency corresponding to the $\gamma_i$ (equations \ref{gi1} 
and \ref{gi2}). $\nu_a$ is the synchrotron {\sl self-absorption} frequency and $\nu_c$ is the
{\sl cooling} frequency; both are calculated below.

 $B$ is the magnetic field in the forward-shock, parametrized by the fraction $b$ of the 
post-shock energy density $u' = \Gamma n' m_p c^2$ that it contains ($u'_B = B^2/8 \pi$), 
with $n' = 4\Gamma n$ the post-shock proton density and $n$ the external medium proton density 
at the location $R(t)$ of the forward-shock. Thus
\begin{equation}
 B = (32\pi b \Gamma^2 n m_p c^2)^{1/2} = 39\,(n b)^{1/2} \Gamma_2 \; {\rm G}
\label{B}
\end{equation}
The synchrotron and inverse-Compton emissions depend on $n$ and $b$ only through the product $nb$. 

 The continuous creation of pairs in the shocked medium, with the power-law distribution given
in equation (\ref{plaw}), and their radiative cooling leads to an effective pair distribution
with energy
\begin{equation}
 \frac{dN}{d\gamma} \propto \left\{ \begin{array}{ll} 
     \gamma^{-2} & (\gamma_p < \gamma < \gamma_b) \\
     \gamma^{-3} & (\gamma_b < \gamma) \\
      \end{array} \right.
\label{dndg}
\end{equation}
with
\begin{equation}
   \gamma_p \equiv \min(\gamma_i,\gamma_c) \;,\; \gamma_b \equiv  \max(\gamma_i,\gamma_c)
\end{equation}
where $\gamma_c$ is the pair {\sl cooling} Lorentz factor, defined as the energy to which a pair
cools on the dynamical timescale $t'_d=R/(c\Gamma) = (4/3) \Gamma [t/(z+1)]$. The pair number $N$
and minimum Lorentz factor $\gamma_i$ change on a dynamical timescale.

 Pairs created with an energy $\gamma m_e c^2$ in the shock-frame cool radiatively at power
$P' (\gamma) = (4/3)\sigma_e c (B^2/8 \pi) (Y+1) \gamma^2 \equiv c_r \gamma^2 $, where $Y$ is 
the Compton parameter (the inverse-Compton to synchrotron power ratio). Integrating the equation 
for pair cooling, $c_r \gamma^2 = - d(\gamma m_e c^2)/dt'$, one obtains that a pair of high 
initial energy reaches a random Lorentz factor $\gamma (t') = m_e c^2/c_r t'$ after a time $t'$ 
since its creation. 
Therefore, the cooling Lorentz factor is $\gamma_c \equiv \gamma (t'_d) = m_e c^2/c_r t'_d$ 
\begin{equation}
 \gamma_c = \frac{9\pi}{2} (z+1) \frac{m_e c}{\sigma_e} \frac{1}{\Gamma t B^2 (Y+1)}
\end{equation}
\begin{equation}
  \gamma_c = \frac{1150\,\z}{(nb) t_1 \Gamma_2^3} \frac{1}{Y+\Theta (\gamma_c-\gamma_a)} 
\label{gc}
\end{equation}
where $\Theta (x) = 0$ for $x < 0$, $\Theta (x) = 1$ for $x > 0$, and $\gamma_a$ is the random Lorentz
factor of the pairs that radiate at the synchrotron self-absorption frequency $\nu_a$ (see below).
Equation (\ref{gc}) approximates the effect of synchrotron self-absorption by switching
off synchrotron cooling when $\gamma_c$ has decreased to $\gamma_a$. After that, the radiative 
cooling continues only through inverse-Compton scatterings. 
This means that, if $\gamma_c$ calculated from equation (\ref{gc}) for $Y < 1 $ and $\Theta = 0$
is larger than $\gamma_a$, then the correct cooling Lorentz factor is $\gamma_c \siml \gamma_a$.

 From equation (6.53) of Rybicki \& Lightman (1979) for the synchrotron self-absorption coefficient 
for a power-law distribution of particles, it can be shown that the pair optical-thickness to 
{\sl self-absorption} at the peak energy $\nu_p = \min(\nu_i,\nu_c)$ of the intrinsic synchrotron 
spectrum $F_\nu$ is
\begin{equation}
 \tau_p = \frac{5e \tpm}{\sigma_e B\gamma_p^5} = 3.6 \times 10^{15} \frac{\tpm}{B_0\gamma_p^5}
\label{taua}
\end{equation}
where
\begin{equation}
 \tpm = \frac{\sigma_e N}{4\pi R^2} 
\label{taupm}
\end{equation}
is the pair optical-thickness to photon {\sl scattering}. For the pair distribution given in equation 
(\ref{dndg}), the optical-thickness to self-absorption at frequency $\nu$ is 
\begin{equation}
\tau_a (\nu) = \tau_p \left\{ \begin{array}{ll}
 \displaystyle{ \left( \frac{\nu_p}{\nu} \right)^{5/3} } & (\nu < \nu_p) \\
 \displaystyle{ \left( \frac{\nu_p}{\nu} \right)^3 } & (\nu_p < \nu < \nu_b) \\
 \displaystyle{ \left( \frac{\nu_p}{\nu_b} \right)^3 \left( \frac{\nu_b}{\nu} \right)^{7/2} } &
    (\nu_b < \nu) \\   \end{array} \right.
\label{taua1}
\end{equation}
where $\nu_b = \max (\nu_p,\nu_c)$.
From here, the self-absorption frequency $\nu_a$ defined by $\tau_a (\nu_a) = 1$ has a corresponding 
pair Lorentz factor $\gamma_a$ given by
\begin{equation}
 \gamma_a = \left\{ \begin{array}{ll}
    \gamma_p \tau_p^{3/10} & (\gamma_a < \gamma_p) \\
    \gamma_p \tau_p^{1/6} & (\gamma_p < \gamma_a < \gamma_b) \\
    \left( \gamma_p^6 \gamma_b \tau_p \right)^{1/7} & (\gamma_b < \gamma_a) \\
      \end{array} \right.
\label{ga}
\end{equation}

 The Compton parameter $Y$ is the ratio of the pairs energy output in inverse-Compton emission 
to that in synchrotron, hence $Y = P_{ic}/P_{sy} = u'_{sy}/u'_B$, where $u'_{sy}$ is the energy 
density of the synchrotron photons received by a scattering lepton. Synchrotron self-absorption 
reduces $u'_{sy}$ because photons of energy less than $h\nu_a$ are absorbed before being scattered.
However, that reduction is not substantial for the particle distribution given in equation 
(\ref{dndg}) because most of the synchrotron energetic output is at frequencies above $\nu_a$, 
for which the pairs medium is transparent (to self-absorption): 
$i)$ for $\nu_a < \nu_{max}$ with $\nu_{max} \equiv \max(\nu_i,\nu_c)$, the synchrotron output 
above $\nu_a$ is $\nu F_\nu \propto \nu^{1/2}$ at $\nu < \nu_{max}$ and $\nu F_\nu \propto \nu^0$ 
at $\nu_{max} < \nu$, thus all the synchrotron output is above $\nu_a$; 
$ii)$ for $\nu_{max} < \nu_a$, the synchrotron output is $\nu F_\nu \propto \nu^{7/2}$ 
for $\nu_{max} < \nu < \nu_a$ and $\nu F_\nu \propto \nu^0$ at $\nu_a < \nu$, 
hence the reduction of the synchrotron output due to self-absorption is a factor 
$\ln (\nu_{br}/\nu_a)/\ln (\nu_{br}/\nu_{max})$ with $\nu_{br}$ the high-energy end of 
the otherwise diverging $F_\nu (>\nu_{max}) \propto \nu^{-1}$ synchrotron spectrum, thus the
reduction is of order unity for $\nu_{br} \gg \nu_a$.

 For a single photon scattering, the ratio $u'_{sy}/u'_B$ is the product of $(4/3)\overline{\gamma^2}$ 
(the average increase in photon energy due to scattering) and the fraction $\min(\tpm,1)$ of 
photons that are upscattered. $\overline{\gamma^2} \simeq \gamma_i \gamma_c$ for the pair distribution 
of equation (\ref{dndg}), thus
\begin{equation}
 Y_1 = \frac{4}{3} \gamma_i \gamma_c  \min(\tpm, 1)
\label{Y}
\end{equation}
for the first scattering. 

 In the lab-frame, the relative velocity between a photon (moving at $c$) and the pair-front 
(moving at Lorentz factor $\Gamma$) is $v_r = c/(2\Gamma^2)$, thus the photon crosses the
pair-front of geometrical thickness $\Delta = R/(2 \Gamma^2)$ in a time $t_+ = \Delta/v_r = R/c = t_d$. 
This means that, for $\tpm < 1$, only the first scattering takes place within a dynamical 
timescale $t_d$, and higher order scatterings occur on a longer timescale.
Because the effective pair distribution with energy is that resulting from pair creation 
and cooling over one $t_d$, higher order scatterings (taking longer than $t_d$) are ignored 
for $\tpm < 1$. However, higher order scatterings should be considered for $\tpm > 1$ because,
in that case, the time between scatterings is less than $t_d$.

\vspace*{2mm}
\section{Regions in the ($\nnbb-\Gamma$) parameter space}
\label{regions}

 To obtain the pair synchrotron self-Compton emission at some observing frequency $\nu$, one must 
calculate first the spectral breaks of the previous section. The calculation of the injection break 
$\nu_i$ is trivial, the analytical expression of the cooling break $\nu_c$ depends on whether 
the Compton parameter $Y$ is below or above unity, that of $Y$ depends on whether the pair 
optical thickness $\tpm$ is above/below unity. In general, the self-absorption break $\nu_a$ 
is not needed for the synchrotron emission, as the optical thickness to self-absorption 
$\tau_a(\nu)$ suffices. However, the location of $\nu_a$ relative to $\nu_c$ is useful for the 
calculation of $\nu_c$ when $Y < 1$ (see equation \ref{gc}), and is required for determining 
the upscattered absorption break of the inverse-Compton spectrum. 

 Given the observables $\Phi$, $t$, and $z$, all quantities needed depend on only two parameters:
$nb$ and $\Gamma$. The conditions $\tpm = 1$, $Y=1$, and $\gamma_c = 1$ define {\sl lines} in this
$nb-\Gamma$ plane while the equality of two spectral breaks defines {\sl boundaries}. Expressions 
for these lines and boundaries are derived below and are useful for a correct calculation of the 
synchrotron/inverse-Compton spectral breaks and of the peak flux for each spectral component.

 The first separation of the $nb-\Gamma$ plane is provided by $\Gamma = \Gamma_c$ (equation \ref{Gc}), 
across which $N$ changes from equation (\ref{N2}) to (\ref{N1}) and $\gamma_i$ from equation 
(\ref{gi1}) to (\ref{gi2}). A similarly simple separation is provided by the condition $\tpm = 1$. 
Using equations (\ref{R}), (\ref{N2}), (\ref{N1}), and (\ref{taupm}), the pair optical-thickness to 
photon scattering is
\begin{equation}
  \tpm = \left\{ \begin{array}{ll}
     3.3\, \z^5 \phi5 t_1^{-2} \Gamma_2^{-5} & \Gamma < \Gamma_c \\
     4.0\, \z^{10} \phi5^2 t_1^{-4} \Gamma_2^{-10} & \Gamma_c \ll \Gamma \\
     \end{array} \right.
\label{tpm0}
\end{equation}
which, after using equation (\ref{Gc}), can be written as 
\begin{equation}
 \tpm = \frac{\sigma_e}{\bar{\sigma}_{\gamma\gamma}} \left\{ \begin{array}{ll} 
    ( \Gamma_c/\Gamma )^{5} & \Gamma < \Gamma_c \\
    2( \Gamma_c/\Gamma )^{10} & \Gamma_c \ll \Gamma \\ \end{array} \right.
\label{tpm}
\end{equation}
This implies that $\tpm < 1$ for $\Gamma > \Gamma_\tau$ where
\begin{equation}
  \Gamma_\tau \equiv \left( \frac{2\sigma_e}{\bar{\sigma}_{\gamma\gamma}} \right)^{1/10} \Gamma_c = 
    1.27\; \Gamma_c = 115\, \z \frac{\phi5^{0.2}}{t_1^{0.4}} 
\end{equation}

 The extrapolations of the $\Gamma < \Gamma_c$ and $\Gamma \gg \Gamma_c$ branches of equations 
(\ref{gi2}) and (\ref{tpm}) intersect at $\tGc \equiv 2^{1/5} \Gamma_c$ for both 
equations. For that reason, we approximate them by
\begin{equation}
 \tpm = \frac{\sigma_e}{2\bar{\sigma}_{\gamma\gamma}} \left\{ \begin{array}{ll} 
    (\tGc/\Gamma )^{5} & \Gamma < \tGc \\ (\tGc/\Gamma )^{10} & \tGc < \Gamma \\ \end{array} \right.
\label{tpm1}
\end{equation}
\begin{equation}
 \gamma_i = \left\{ \begin{array}{ll} 2 & (\Gamma < \tilde{\Gamma}_c) \\
    2 (\Gamma/\tGc )^5 & (\tGc < \Gamma) \\ \end{array} \right.
    \quad \tGc \equiv 104\, \z \frac{\phi5^{0.2}}{t_1^{0.4}} 
\label{gi3}
\end{equation}

 The range $\tGc < \Gamma < \Gamma_\tau$, where $\gamma_i > 2$ and $\tpm > 1$, 
is narrow, its 10 percent extent in $\Gamma$ corresponding to a 30 percent increase in observer 
time, for a shock decelerating as $\Gamma \propto t^{-3/8}$ (homogeneous external medium),
i.e. it lasts less than one dynamical timescale. 

\begin{figure*}
\vspace*{-5mm}
\centerline{\psfig{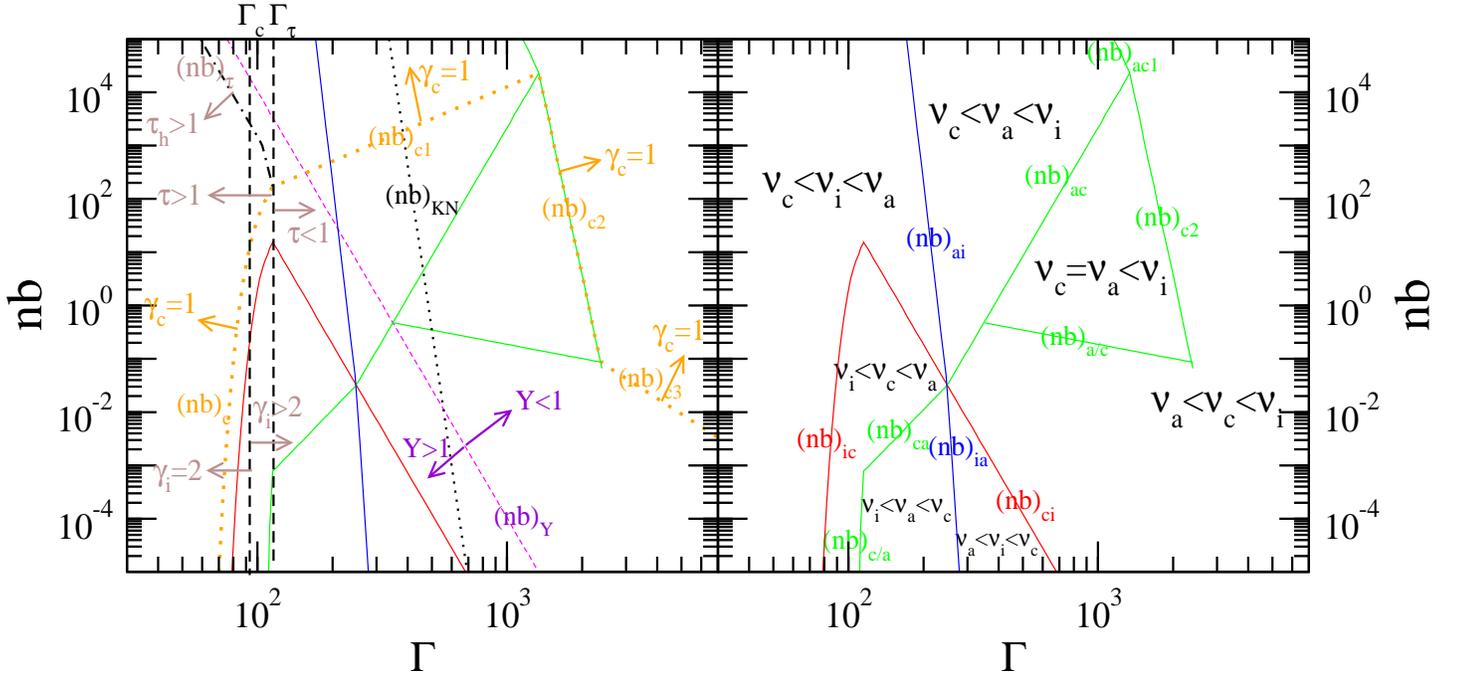}}
\figcaption{ {\sl Boundaries} (solid curves, labeled in right panel) and {\sl Lines} (dotted and 
   dashed curves, labeled in the left panel) in the $nb - \Gamma$ plane, and regions corresponding   
   to all possible orderings of spectral breaks (separated by the boundaries identified in right panel),
   for the synchrotron emission from pairs formed at $z=2$ from high-energy photons of fluence 
   $\Phi (>100\,{\rm MeV})=10^{-5} \ergcm2$ at observer time $t = 10$ s. 
   $\Gamma$ is the Lorentz factor of the pair-source at that time, and $nb$ is its magnetic field parameter. 
   The boundaries intersect at the "triple point" $(nb,\Gamma) = (0.035,250)$, where all three breaks 
   are equal. 
   For $\Gamma < \Gamma_\tau$, the pairs are optical thick to photon scattering $\tau > 1$, but 
   for $nb > (nb)_\tau$, the hot pairs ($\gamma > 1$) are optically thin. } 
\end{figure*}

\vspace*{2mm}
\subsection{Optically-thin pairs: $\Gamma > \Gamma_\tau$}

  Equation (\ref{Y}) is simply
\begin{equation}
 Y = \frac{4}{3} \gamma_i \gamma_c \tpm  (= Y_1)
\label{Y1}
\end{equation}
retaining only the first scattering, together with equation (\ref{gc}) leading to
\begin{equation}
 Y(Y+1) = \frac{(nb)_Y}{nb} \;, \quad (nb)_Y \equiv 10^4\, \z^6 \frac{\phi5}{t_1^3} \Gamma_2^{-8}
\label{nbY}
\end{equation}
The line $nb = (nb)_Y$ separates the $nb-\Gamma$ plane in two regions, with $Y < 1$ for
$nb > (nb)_Y$ and $Y > 1$ for $nb < (nb)_Y$ (see Figure 1).

\subsubsection{\boldmath $nb < (nb)_Y$ \unboldmath (iC cooling)} 

 In this case $Y > 1$, and equations (\ref{gc}) and (\ref{nbY}) yield
\begin{equation}
 \gamma_c = 11\, \z^{-2} \left( \frac{t_1 \Gamma_2^2}{(nb) \phi5} \right)^{1/2} \;, \quad
 Y \simeq \left( \frac{(nb)_Y}{nb} \right)^{1/2} 
\label{Y2}
\end{equation}
\centerline{ $(\tau < 1, Y > 1, \gamma_c > 1)$ } 

 From equations (\ref{gi2}) and (\ref{Y2}), the condition $\gamma_i = \gamma_c$ defines the boundary
\begin{equation}
 (nb)_{ci} \equiv 47\, \z^6 \frac{\phi5}{t_1^3} \Gamma_2^{-8}
\label{nbci}
\end{equation}
so that $\gamma_i < \gamma_c$ for $nb < (nb)_{ci}$. From equations (\ref{nbY}) and (\ref{nbci}),
it follows that 
\begin{equation}
 \frac{(nb)_{ci}}{(nb)_Y}=\frac{9}{64}\left(\frac{\bar{\sigma}_{\gamma\gamma}}{\sigma_e}\right)^2 = 
     \frac{1}{219}
\end{equation}
thus $(nb)_{ci}$ is just a shift of $(nb)_Y$.

 For $nb < (nb)_{ci}$, we have $\gamma_i < \gamma_c$ and the condition $\gamma_a = \gamma_i$ 
is equivalent to $\tau_p \equiv \tau_a (\gamma_i) = 1$ which, after using equations (\ref{gi2}), 
(\ref{taua}), and (\ref{tpm0}), is satisfied on the boundary
\begin{equation} 
 (nb)_{ia} \equiv 0.014\, \z^{70} \frac{\phi5^{14}}{t_1^{28}} \Gamma_{2.4}^{-72}
\end{equation}
with $\gamma_i < \gamma_a$ for $nb < (nb)_{ia}$.
In this regime, $\gamma_a = \gamma_i \tau_p^{1/6}$; using equations (\ref{gi2}), (\ref{taua}), 
(\ref{tpm0}), and (\ref{Y2}), $\gamma_a = \gamma_c$ is satisfied on the boundary 
\begin{equation} 
 (nb)_{ca} \equiv 0.011 \z^{-6.8} \frac{t_1^2}{\phi5^{1.6}} \Gamma_{2.3}^{4.8}
\end{equation}
with $\gamma_a < \gamma_c$ if $nb < (nb)_{ca}$. 

 For $nb > (nb)_{ci}$, we have $\gamma_c < \gamma_i$ and the $\gamma_a = \gamma_c$ boundary 
is defined by $\tau_p \equiv \tau_a (\gamma_c) = 1$ which, together with equations (\ref{taua}), 
(\ref{tpm0}), and (\ref{Y2}), leads to
\begin{equation} 
 (nb)_{ac} \equiv 8.8\, \z^{-10} \frac{t_1^{3.25}}{\phi5^{2.25}} \Gamma_{2.7}^8
\end{equation}
with $\gamma_a > \gamma_c$ for $nb > (nb)_{ac}$. In this case, $\gamma_a = \gamma_c \tau_p^{1/6}$;
using equations (\ref{gi2}), (\ref{taua}), (\ref{tpm0}), and (\ref{Y2}), it follows that 
$\gamma_a = \gamma_i$ on the boundary 
\begin{equation} 
 (nb)_{ai} \equiv 200\, \z^{38} \frac{\phi5^{7.5}}{t_1^{15.5}} \Gamma_{2.3}^{-40}
\label{ai}
\end{equation}
such that $\gamma_i < \gamma_a$ if $nb < (nb)_{ai}$.

\vspace*{3mm}
\subsubsection{\boldmath $(nb)_Y < nb < (nb)_{a/c}$ \unboldmath (sy cooling)}

 In this case $Y < 1$; for $\gamma_a < \gamma_c$, pairs cool mostly through synchrotron
emission and equations (\ref{gc}) and (\ref{nbY}) yield
\begin{equation}
 \gamma_c \simeq \frac{1.15\,\z}{(nb) t_1 \Gamma_3^3} \;,\; Y = \frac{(nb)_Y}{nb} \quad 
    (\tau < 1, Y < 1, \gamma_c > \gamma_a)
\label{Y3}
\end{equation}
Because $(nb)_Y > (nb)_{ci}$, we have $\gamma_c < \gamma_i$, and the condition $\gamma_a = \gamma_c$ 
is equivalent to $\tau_p \equiv \tau_a (\gamma_c) = 1$ which, using equations (\ref{taua}), 
(\ref{tpm0}), and (\ref{Y3}), is satisfied on the boundary
\begin{equation} 
 (nb)_{a/c} \equiv 0.19\, \frac{\z^{-1.11}}{t_1^{0.22} \phi5^{0.44}} \Gamma_3^{-0.89}
\end{equation}

\subsubsection{\boldmath $ nb > \max[(nb)_Y, (nb)_{a/c}]$ \unboldmath}

 For $nb > (nb)_{a/c}$, the $\gamma_c < \gamma_a$ regime occurs; without synchrotron
cooling, $\gamma_c \propto Y^{-1}$, and the expressions for $Y$ and $\gamma_c$ are the 
same as in equation (\ref{Y2}) for $Y > 1$, although $Y < 1$ now. 
Consequently, the $\gamma_a = \gamma_i$ and $\gamma_a = \gamma_c$ boundaries are the same 
as for $Y>1$: $(nb)_{ai}$ and $(nb)_{ac}$, respectively. 

 In contrast with that case, 
a new region appears now, defined by $(nb)_{a/c} < nb < (nb)_{ac}$, for which $\gamma_c$
of equation (\ref{Y2}) does not satisfy the $\gamma_c < \gamma_a$ condition. This is the
case where inverse-Compton cooling, operating alone after the epoch $t'_{ac} < t'_d$ when 
the cooled pair Lorentz factor $\gamma(t'_{ac})$ has reached $\gamma_a(t'_{ac})$, does not 
decrease $\gamma(t')$ significantly until $t'_d$. In this case, we impose $\gamma_c \equiv 
\gamma (t'_d) \simeq \gamma_a$, thus 
\begin{equation} 
 \gamma_c = \gamma_a = \left( \frac{5e \tpm}{\sigma_e B} \right)^{1/5} = 
   5.2\, \z^2 \left( \frac{\phi5^2}{(nb)^{1/2} t_1^4 \Gamma_3^{11}} \right)^{1/5}
\label{gc1}
\end{equation}
\centerline{ $(\tpm < 1, Y < 1, \gamma_c=\gamma_a)$ }\\

\subsubsection{\boldmath $nb > (nb)_{c1}$ \unboldmath or 
               \boldmath $nb > \max[(nb)_{c2},(nb)_{c3}]$ \unboldmath} 
\centerline{\bf $(\gamma_c=1)$ }

 In some of the regions identified above, radiative cooling can be strong enough that 
$\gamma (t') = 1$ at some time $t' < t'_d$. In this case, $\gamma_c \equiv \gamma (t'_d) = 1$, 
and $\gamma_a$ and $Y$ are those given by equations (\ref{ga}) and (\ref{Y1}) with $\gamma_c = 1$. 
 From equations (\ref{Y2}), (\ref{Y3}), and (\ref{gc1}), the condition $\gamma_c = 1$ defines 
three lines:
\begin{equation} 
 (nb)_{c1} \equiv 1270\, \z^{-4} \frac{t_1}{\phi5^2} \Gamma_{2.5}^2
\end{equation}
\begin{equation} 
 (nb)_{c2} \equiv 3.6\, \z^{20} \frac{\phi5^4}{t_1^8} \Gamma_{3.3}^{-22}
\label{c2}
\end{equation}
\begin{equation} 
 (nb)_{c3} \equiv 0.043\, \frac{\z}{t_1} \Gamma_{3.5}^{-3}
\end{equation}
such that $\gamma_c = 1$ if $nb > (nb)_{c1}$ or $nb > \max[(nb)_{c2},(nb)_{c3}]$. 

 The pair cooling to $\gamma_c=1$ means that a pair of high initial energy loses radiatively 
all its energy over a dynamical timescale. From the equation for radiative cooling, 
$\gamma (t') = m_e c^2/c_r t'$, it follows that the time for complete cooling is 
$t'_c(\gamma \sim 1) \equiv m_e c^2/c_r$. When $t'_c (\gamma \sim 1) < t'_d$, a fraction 
$f_h = t'_c (\gamma \sim 1)/t'_d$ of the pairs created in the last dynamical timescale 
are hot ($\gamma > 1$) and radiate synchrotron and upscatter/absorb that emission, and 
a fraction $1 - f_h $ are cold ($\gamma = 1$), scatter photons without a significant 
change in energy and absorb the emission below their characteristic synchrotron frequency. 
Noting that $t'_c(\gamma \sim 1)/t'_d = m_e c^2/(c_r t'_d) = \gamma_c^{({\rm eq}. \ref{gc})}$, 
with $\gamma_c^{({\rm eq}. \ref{gc})}$ the cooling Lorentz factor of equation (\ref{gc}), 
we find that
\begin{equation} 
 f_h = \gamma_c^{(eq \ref{gc})} = \frac{0.144\,\z}{(nb)_3 t_1 \Gamma^3_{2.3}} 
  \frac{1}{Y+\Theta (1-\gamma_a)} 
\label{fh}
\end{equation}

 The parameters of interest -- $\gamma_a,Y$ and peak flux $F_p \propto N$ -- 
are those of equations (\ref{ga}) and (\ref{Y}) with $\gamma_c = 1$ and for a number of 
hot pairs $N_h$ or optical thickness $\tau_h$ satisfying $N_h/N = \tau_h/\tpm = f_h$. 

 A first consequence of $\tau_h = f_h \tpm = \tpm \gamma_c^{({\rm eq}. \ref{gc})}$ for 
$\gamma_c = 1$ is that $Y = (4/3)\gamma_i \tau_h = (4/3)\gamma_i \gamma_c^{(eq \ref{gc})} \tpm$, 
thus the Compton parameter given in equation (\ref{Y1}) applies whether pairs cool completely 
($f_h < 1$) or not ($f_h = 1$) during one dynamical timescale. That entails the conclusions 
that, for $\gamma_c = 1$, $Y=1$ on the $(nb)_Y$ line given in equation (\ref{Y2}), and that 
$Y$ is as in equation (\ref{Y2}) for $\gamma_c < \gamma_a$ and as in equation (\ref{Y3}) for 
$\gamma_a < \gamma_c$. 

 Owing to the accumulation of pairs at $\gamma_c = 1$, the synchrotron self-absorption thickness 
satisfies $\tau_a (\nu < \nu_c) \propto N$ (all pairs absorb synchrotron emission below $\nu_c$) 
and $\tau_a (\nu > \nu_c) \propto N_h$ (only hot pairs absorb above $\nu_c$), yielding a 
discontinuity of $\tau_a (\nu)$ at $\nu_c$ and a slight ambiguity on the boundary between 
the $\nu_c < \nu_a < \nu_i$ and $\nu_a < \nu_c < \nu_i$ regions, as following.
The $\gamma_a = \gamma_c$ boundary of the $\gamma_c < \gamma_a$ region satisfies 
$\tau_a(\nu_c+) = (5e \tau_h /\sigma_e B)^{1/6} = 1$, with $\tau_h = f_h \tpm \propto \tpm/Y$.
Using equations (\ref{tpm0}), (\ref{Y2}), and (\ref{fh}), we find that $\gamma_a = \gamma_c$
along the boundary
\begin{equation} 
 (nb)_{ac1} \equiv 4.4 \times 10^5\, \z^8 \frac{\phi5^{1.5}}{t_1^{3.5}} \Gamma_3^{-10}
\end{equation}
The $\gamma_a = \gamma_c$ boundary of the $\gamma_a < \gamma_c$ region satisfies 
$\tau_a(\nu_c-) = (5e \tau /\sigma_e B)^{1/6} = 1$, leading to the same condition as for
$\gamma_c = 1$ for the $\nu_a = \nu_c < \nu_i$ region (see Figure 1), therefore the $(nb)_{c2}$ 
line of equation (\ref{c2}) is also a $\gamma_a = \gamma_c$ boundary. 

 For $\gamma_c < \gamma_a < \gamma_i$, equations (\ref{taua}) and (\ref{ga}) give 
$\gamma_a = \gamma_c [\tau_a(\nu_c+)]^{1/6} = (5e \tau_h/\sigma_e B)^{1/6}$; then,
$\tau_h = \tpm \gamma_c^{({\rm eq}. \ref{gc})}$ implies that $\gamma_a = \gamma_i$ along the 
$(nb)_{ai}$ boundary given in equation (\ref{ai}), extending into the $\gamma_c =1$ region.

\subsubsection{Klein-Nishina scattering} 

 The Compton parameter $Y$ of equation (\ref{Y1}) is for inverse-Compton scatterings 
in the Thomson regime. The pair synchrotron flux is $\nu F_\nu \propto \nu^{1/2}$ for 
$\nu_p < \nu < \nu_b$ and flat above $\nu_b$ for a $C_\veps \propto \veps^{-2}$ spectrum
of the high-energy photons that form pairs. However, for a typical LAT spectrum, which
is slightly softer than $\veps^{-2}$, the synchrotron flux $\nu F_\nu$ peaks at $\nu_b$. 
For $nb > (nb)_{ci}$, we have $\nu_c < \nu_i$, hence $\nu F_\nu$ peaks at $\nu_i$.
Pairs of energy $\gamma$ scatter synchrotron photons at the $\nu_i$ peak of $\nu F_\nu$
in the Klein-Nishina (KN) regime if $\gamma (h \nu'_i) = \gamma (eh/2 m_e c) \gamma_i^2 B >
m_e c^2$. Staying in the $\gamma_c < \gamma_i$ case, the KN regime will reduce the Compton 
parameter significantly if the $\gamma_i$ pairs satisfy the above inequality, i.e. if
$\gamma_i > \gamma_{KN} \equiv (2 m_e^2 c^3/heB)^{1/3}$. Using equations (\ref{gi2}) and 
(\ref{B}), the condition $\gamma_i > \gamma_{KN}$ becomes $nb > (nb)_{KN}$ where 
\begin{equation}
 (nb)_{KN} \equiv 0.29\, \z^{30} \frac{\phi5^6}{t_1^{12}} \Gamma_{2.7}^{-32}
\label{nkb}
\end{equation}

 For $nb > 10^{-5}$, we have $(nb)_{KN} > (nb)_{ci}$ (Figure 1), thus $nb > (nb)_{KN}$
implies $nb > (nb)_{ci}$ and the derivation of equation (\ref{nkb}) is self-consistent. 
The KN scattering effect on the calculation of the Compton parameter of equation (\ref{Y1}) 
is as following.
For $nb < (nb)_{KN}$, most pairs, being between $\gamma_c$ and $\gamma_i$, scatter the 
synchrotron input in Thomson regime and the KN effect is negligible.
For $nb > (nb)_{KN}$, pairs above an energy $\tilde{\gamma} \equiv \gamma_{KN}^3/\gamma_i^2$ 
with $\gamma_c < \tilde{\gamma} < \gamma_i$ scatter synchrotron photons in the KN regime
and the Compton $Y$ is reduced; however, $Y < 1$ is very likely because $nb > (nb)_{KN}$ 
implies $nb > (nb)_Y$ (Figure 1), hence that reduction of $Y$ by the KN effect is 
largely inconsequential.

 For $nb < 10^{-5}$, we have $\gamma_{KN} < \gamma_i < \gamma_c$; in this case, all pairs
upscatter in the KN regime the $\nu_i$ synchrotron photons and also the $\nu_c$ synchrotron
photons (where $\nu F_\nu$ peaks), thus equation (\ref{Y1}) significantly overestimates the 
true Compton parameter, leading to an overestimation of the inverse-Compton flux and an 
underestimation of $\gamma_c \propto Y^{-1}$. The latter leaves unchanged the synchrotron 
flux below $\nu_c$ but underestimates the synchrotron flux above $\nu_c$. For the rest
of this paper, we will avoid the $nb < 10^{-5}$ region, so that the KN effect can be ignored.

\vspace*{2mm}
\subsection{Optically-thick pairs: $\Gamma < \Gamma_\tau$}

 Over one dynamical timescale, a photon undergoes $\tpm$ scatterings and diffuses a distance 
$L'_s = \sqrt{\tpm} l'_s$ in the pair-shell comoving frame, where $l'_s = \Delta'/\tpm$ is the 
mean free-path between scatterings. Thus, over one dynamical timescale, the observer receives 
photons from a geometrical depth $L'_s = \Delta'/\sqrt{\tpm}$, corresponding to a scattering optical 
depth $\sqrt{\tpm}$.

 A synchrotron photon undergoes $\tpm^2$ scatterings before escaping the pair medium. 
 In the lab-frame, if the pair medium were stationary, the photon mean free-path between scatterings 
would be $l_s = \Delta/\tpm$, with $\Delta = R/(2 \Gamma^2)$ the geometrical thickness of the pair 
front. Because the pair medium moves at Lorentz factor $\Gamma$, the lab-frame photon-pair
relative velocity is $v_r = c/(2\Gamma^2)$, hence each scattering takes a time $t_s = 2\Gamma^2 l_s/c=
R/(c\tpm) = t_d/\tpm$. A photon starting from a geometrical depth $x' = \tau_\o (\Delta'/\tpm)$, 
corresponding to a scattering optical depth $\tau_\o$, will diffuse through the pair medium a distance 
$L'_d = \sqrt{n_s} l'_s$ (measured relative to the forward-shock) after $n_s$ scatterings and will exit 
the pair medium in the direction toward the observer when $L'_d = x'$, i.e. after $n_s = (x'/l'_s)^2=
\tau_\o^2$ scatterings, in a lab-frame time $t_{esc} = n_s t_s = \tau_\o^2 t_d/\tpm$.
Therefore, only photons up to an optical thickness depth $\sqrt{\tpm}$ cross the pair shell
within one dynamical timescale, and these photons undergo up to $\tpm$ scatterings.

 Because we approximate the effective pair distribution with energy as that established by injection 
and cooling over $t_d$, we will consider only $\tpm$ scatterings for the calculation of inverse-Compton
parameter, cooling, and emission. 
Hence, the pair cooling $\gamma_c$ is that of equation (\ref{gc}) with the Compton parameter 
corresponding to $\tpm$ scatterings: $Y = Y_1 + ... + Y_{\tpm}$. As long as scatterings 
occur in the Thomson regime, the Compton parameter of $i$-th scattering is $Y_i = Y_1^i$,
where $Y_1 = (4/3) \gamma_i \gamma_c$ is the Compton parameter of the first scattering
for $\gamma_c > 1$. Because $Y_1 > 1$, we can approximate $Y_1 + ... + Y_{\tpm} =
(Y_1^{\tpm +1} -1)/(Y_1-1) \simeq Y_1^{\tpm}$, thus
\begin{equation}
 Y = Y_1^\tpm \;,\;  Y_1 = \frac{4}{3} \gamma_i \gamma_c  \quad (\tau > 1, \gamma_c > 1)
\end{equation} 
Adding equation (\ref{gc}) with $Y > 1$, we find that
\begin{equation}
 \gamma_c = \left[ \left( \frac{3}{4\gamma_i} \right)^\tpm 
         \frac{1150\, \z}{(nb) t_1 \Gamma_2^3} \right]^{1/(\tpm+1)} \quad (\tau > 1, \gamma_c > 1)
\label{gc2}
\end{equation} 

 After $i$ scatterings, a synchrotron photon of initial energy $h \nu$ reaches an energy
$(\overline{\gamma^2})^i h\nu$, where $\overline{\gamma^2} = \gamma_i \gamma_c$ is the average 
$\gamma^2$ for pairs. The $i$-th scattering occurs at the end of the Thomson regime if, 
in the pair frame, the $(i-1)$-th scattered photon is below $m_e c^2$ (Thomson scattering) 
and the $i$-th scattered photon is above $m_e c^2$ (Klein-Nishina scattering): 
$\bar{\gamma} (\overline{\gamma^2})^{i-1} h\nu < m_e c^2 < \bar{\gamma} 
(\overline{\gamma^2})^i h\nu$, where $\bar{\gamma} =(\overline{\gamma^2})^{1/2} = 
(\gamma_i \gamma_c)^{1/2}$. Thus the $i$-th scattering is borderline in Thomson regime 
if $(\gamma_i \gamma_c)^{i-1/2} h\nu < m_e c^2 < (\gamma_i \gamma_c)^{i+1/2} h\nu$,
and we approximate the number of Thomson scatterings $n_T$ by 
$(\gamma_i \gamma_c)^{n_T} h\nu = m_e c^2$. 

 The synchrotron photon of energy $h \nu$ considered here should be that where most of the 
synchrotron output $\nu F_\nu$ lies, which is $\max (\nu_i, \nu_c)$, leading to
\begin{equation}
 \max(\gamma_i^2,\gamma_c^2) (\gamma_i \gamma_c)^{n_T} = \frac{2 m_e^2 c^3}{heB} = 
          \frac{3.6 \times 10^{11}}{(nb)^{1/2} \Gamma_2}
\label{heb}
\end{equation}
Substituting here $\gamma_c$ from equation (\ref{gc2}) and using equations (\ref{tpm1}) 
and (\ref{gi3}), one can determine the $nb-\Gamma$ region where $\tpm < n_T$, i.e. for 
which all $\tpm$ scatterings occurring within one dynamical timescale are in the Thomson 
regime. The result is that, for the $nb$ range shown in Figure 1, all $\tpm$ upscatterings 
occur in the Thomson regime, and that result can be illustrated easier if we use the more 
stringent condition that $\nu_c$ photons are upscattered by $\gamma_c$ pairs. Then, from 
equation (\ref{heb}) with $\gamma_i$ substituted by $\gamma_c$ and with $\gamma_c$ from
equation (\ref{gc2}), it follows that $\tpm < n_T$ is equivalent to $nb > (nb)_{kn}$ with
\begin{equation}
 (nb)_{kn} \equiv 2.4 \times 10^{-4} \left[ \left( \frac{3}{4\gamma_i} \right)^{\tpm} 
         \frac{\z}{t_1 \Gamma_2^{2.5}} \right]^{4/3}
\end{equation}
Taking into account the dependence on $\Gamma$ of $\tpm$ and $\gamma_i$, we find that $(nb)_{kn}$
above rises with $\Gamma$, reaching a maximum value at $\Gamma_\tau$, where $\tpm = 1$ and 
$\gamma_i = 3.3$, hence that maximum value is $(nb)_{kn}(\Gamma_\tau) \simeq 2.10^{-5}$.
Thus, the $nb$ range shown in Figure 1 satisfies $nb > (nb)_{kn}$ and all $\tpm$ upscatterings 
occur in the Thomson regime.

\vspace*{2mm}
\subsubsection{\boldmath $\gamma_c > 1$ \unboldmath }

 From equation (\ref{gc2}), it follows that $\gamma_c = \gamma_i$ along the boundary
\begin{equation}
 (nb)_{ic} \equiv \left( \frac{3}{4\gamma_i^2} \right)^{\tpm} 
       \frac{1150\, \z}{\gamma_i t_1 \Gamma_2^3}
\end{equation} 
that $\gamma_c = \gamma_a (= \gamma_i \tau_p^{1/6})$ for 
\begin{equation}
 (nb)_{c/a} \equiv \left[ 3.9\, \frac{\z^{-(5\tpm-1)/6}}{(390 \gamma_i)^\tpm} 
     \frac{t_1^{(\tpm-2)/3} \Gamma_2^{\tpm-2}} {\phi5^{(\tpm +1)/6}} \right]^{12/(11-\tpm)}
\end{equation} 
(this boundary exists only for $\tpm < 11$, i.e. $\Gamma > \Gamma_c$), 
and that $\gamma_c = 1$ on
\begin{equation}
 (nb)_c \equiv \left( \frac{3}{4\gamma_i} \right)^{\tpm} \frac{1150\, \z}{t_1 \Gamma_2^3}
    = \gamma_i^{\tpm +1} (nb)_{ic}
\end{equation} 
with $\gamma_i$ given in equation (\ref{gi3}). These results become simpler for $\Gamma < \tGc$,
where $\gamma_i = 2$ and the $(nb)_c$ boundary is just a shift of $(nb)_{ic}$.

\vspace*{2mm}
\subsubsection{\boldmath $\gamma_c = 1$ \unboldmath }

 For $nb > (nb)_c$, only a fraction
\begin{equation} 
 f_h = \frac{1150\,\z}{(nb)t_1 \Gamma_2^3} \frac{1}{Y} \;,\;
  Y=\sum_{i=1}^{\max(1,\tau_h)} Y_1^i \;,\;  Y_1 = \frac{4}{3} \gamma_i \min(1,\tau_h) 
\label{fh1}
\end{equation}
\centerline {$(\tau > 1, \gamma_c=1)$} \\ \\
of all pairs {\sl upscatter} synchrotron photons. This result is equation (\ref{fh}) for 
$\tau_h = f_h \tpm$ upscatterings (a synchrotron photon undergoes $\tpm$ scatterings in one 
dynamical timescale, out of which a fraction $f_h < 1$ are upscatterings by hot pairs) and 
for $\gamma_a > \gamma_c = 1$ (as Figure 1 shows for $\Gamma < \Gamma_\tau$). 
 To find $f_h$, $\tau_h$, and $Y$, equations (\ref{fh1}) with $\tau_h = f_h \tpm$ must be solved.
 
 For $\tau_h > 1$, we have $Y_1>1$ and $Y=\sum_{i=1}^{\tau_h} Y_1^i \simeq Y_1^{\tau_h}$, 
thus $\tau_h$ satisfies
\begin{equation} 
 \left( \frac{4}{3} \gamma_i \right)^{\tau_h} \tau_h  = \frac{1150\,\z \tpm}{(nb)t_1 \Gamma_2^3} 
  \;,\; Y = \left( \frac{4}{3}\gamma_i \right)^{\tau_h} \; (\tau_h > 1, \gamma_c =1)
\label{th}
\end{equation}
From here, $\tau_h$ can be determined numerically, a rough approximation being
$\tau_h \simeq \log [1150\,\z \tpm/(nb)t_1 \Gamma_2^3]/ \log (4 \gamma_i/3)$.

 From equation (\ref{th}), the line corresponding to $\tau_h = 1$ is
\begin{equation} 
 (nb)_\tau  = \frac{870\, \z \tpm}{t_1 \Gamma_2^3 \gamma_i} = \left\{ \begin{array}{ll}
    2200\, \z^{16} \displaystyle{ \frac{\phi5^3}{t_1^7} \Gamma_2^{-18}} & 
                   \tilde{\Gamma}_c < \Gamma < \Gamma_\tau \\
    1470\, \z^6 \displaystyle{ \frac{\phi5}{t_1^3} \Gamma_2^{-8} } & \Gamma < \tilde{\Gamma}_c \\
          \end{array} \right.
\end{equation}
For $nb < (nb)_\tau$, the hot pairs are optically thick ($\tau_h > 1$) to photon scattering 
while for $nb > (nb)_\tau$, they are optically thin ($\tau_h < 1$) despite that $\tpm > 1$. 

 For $\tau_h < 1$, at most one upscattering occurs within a dynamical timescale, thus 
$Y = Y_1 = (4/3)\gamma_i \tau_h$, and  
\begin{equation} 
 \tau_h = \left( \frac{(nb)_\tau}{nb}\right)^{1/2}  \;,\;
   Y = \frac{4}{3} \gamma_i \tau_h \quad  (\tau_h < 1, \gamma_c =1)
\end{equation}
which implies that $Y > 1$ for $nb < (nb)_Y$ with
\begin{equation} 
  (nb)_Y = \left( \frac{4}{3} \gamma_i \right)^2 (nb)_\tau = 
   1540\, \z \frac{\gamma_i \tpm}{t_1 \Gamma_2^3} = (nb)_Y^{({\rm eq}.\ref{nbY})}
\end{equation}
with the last equality resulting from equations (\ref{tpm1}) and (\ref{gi3}).
Thus the $Y=1-(nb)_Y$ line given in equation (\ref{nbY}) for $\Gamma > \Gamma_\tau$ extends
in the $\Gamma < \Gamma_\tau$ region. 
We also note that $Y = (4/3)\gamma_i \tau_h$ implies that, for $\Gamma < \tilde{\Gamma}_c$ 
(i.e. for $\gamma_i = 2$), the $\tau_h = 1$ line is a shift of the $Y=1$ line.

\vspace*{2mm}
\section{Pair emission}
\label{moresyic}

\subsection{Synchrotron} 

 The intrinsic synchrotron spectrum peaks at $\nu_p = \min(\nu_i,\nu_c)$, where the flux 
density is
\begin{equation}
 F_p = \frac{z+1}{4\pi d_l^2} \frac{e^3}{m_e c^2} N_h B \Gamma =
      6.8\, \z^5 \frac{\phi5^2 (nb)^{1/2}}{t_1^2 \Gamma_{2.5}^4} \; {\rm Jy} \;(\Gamma > \tGc)
\label{Fp}
\end{equation}
with $N_h = N$ if $\gamma_c > 1$ and $N_h = f_h N$ if $\gamma_c = 1$. 
 From equations (\ref{nui}), (\ref{B}),(\ref{gi2}), (\ref{Y2}), (\ref{Y3}), the injection and
cooling frequencies scale as
\begin{equation}
 \nu_i \propto n^{1/2} \z^{-11} \Gamma^{12} \Phi^{-2} t^4 \;,\;
 \nu_c \propto \left\{ \begin{array}{ll} \z^{-3} n^{1/2} \Gamma^{-4} t^{-2} & Y < 1 \\
              \z^{-5} n^{-1/2} \Gamma^{4} \Phi^{-1} t & Y > 1 \end{array}  \right.
\label{nuic}
\end{equation}

 For the pair distribution with energy given in equation (\ref{dndg}), the spectrum of the 
{\sl unabsorbed} synchrotron emission is
\begin{equation}
 F_\nu^{(o)} = F_p \left\{ \begin{array}{ll}  
 \displaystyle{ \left(\frac{\nu}{\nu_p}\right)^{1/3} } & (\nu<\nu_p) \\
 \displaystyle{ \left(\frac{\nu_p}{\nu}\right)^{1/2} } & (\nu_p<\nu<\nu_b) \\
 \displaystyle{ \left(\frac{\nu_p}{\nu_b}\right)^{1/2} \frac{\nu_b}{\nu} } & (\nu_b<\nu) \\
   \end{array} \right.
\label{Fnu}
\end{equation}
where $\nu_p = \min(\nu_i,\nu_c)$ and $\nu_b = \max(\nu_i,\nu_c)$.
For a LAT photon spectrum $C_\veps \propto \veps^{-2}$, the corresponding pair synchrotron 
$\nu F_\nu$ spectrum is flat above $\nu_b$, but peaks at $\nu_b$ for softer LAT photon spectra.

 1) $\Gamma > \Gamma_\tau$.  
Then $\tpm < 1$ and the received/emergent synchrotron flux is that of equation (\ref{Fnu}) 
corrected only for self-absorption:
\begin{equation}
 F_\nu = \frac{F_\nu^{(o)}}{\max[1,\tau_a(\nu)]} \quad (\tpm < 1)
\label{Fnu1}
\end{equation}
where $\tau_a (\nu)$ is the synchrotron self-absorption optical thickness (equation \ref{taua1})
at observing frequency $\nu$, for all pairs (if $\gamma_c > 1$ or if $\gamma_c =1$ and $\nu < \nu_c$)
or only for hot pairs (if $\gamma_c =1$ and $\nu > \nu_c$).
Equation (\ref{Fnu1}) simply means that the observer receives emission from the entire pair medium
at frequencies where $\tau_a (\nu) < 1$ and from a layer of optical depth (to self-absorption) unity 
if $\tau_a (\nu) > 1$.
 
 Substituting equations (\ref{Fp}) and (\ref{nuic}) in (\ref{Fnu}), we arrive at the synchrotron 
light-curves listed in Table 1 for $\nu > \nu_a$, $\gamma_c > 1$, and for two types of ambient medium: 
{\sl homogeneous}, for which $n$= const and the shock deceleration is given by $\Gamma^2 n R^3 = 
\Gamma^2 n (\Gamma^2 t)^3=$ const, hence $\Gamma \propto t^{-3/8}$; 
{\sl wind}-like, where $n \propto r^{-2}$, thus $\Gamma \propto t^{-1/4}$, $R \propto t^{1/2}$, and 
$n \propto t^{-1}$. 

\begin{table*}[t]
 \caption{Synchrotron light-curves at frequencies above self-absorption, for optically-thin pairs, 
   for forward-shock (external) electrons (eqs in Appendixes B and C of Panaitescu \& Kumar 2000), 
   and for reverse-shock (ejecta) electrons undergoing adiabatic cooling, i.e. after the reverse shock 
   has crossed the ejecta shell and when there is no further injection of fresh electrons 
  (the $\nu < \nu_c$ flux scalings are inferred from eqs 52 and 53 of Panaitescu \& Kumar 2004, 
  the $\nu_c < \nu$ flux decay is that of the "large-angle emission" derived by Kumar \& Panaitescu 2000). 
  The injected electrons and formed pairs have a power-law distribution with energy of index -2 
  (equation \ref{plaw}). 
  $\Phi$ is the fluence at time $t$ of the high-energy photons that form pairs, $Y$ is the
  Compton parameter, $\nu_i$ and $\nu_c$ are the injection and cooling frequencies.
  The synchrotron spectrum is given in equation (\ref{Fnu}).  }
\vspace*{2mm} 
\begin{tabular}{cccccccccccccc}
  \hline \hline
 type of 
 & $F(\nu<\nu_i<\nu_c)$ & \multicolumn{2}{c}{$F(\nu<\nu_c<\nu_i)$} & $F(\nu_i<\nu<\nu_c)$ & 
   \multicolumn{2}{c}{$F(\nu_c<\nu<\nu_i)$} & \multicolumn{2}{c}{$F(\nu_i,\nu_c<\nu)$} \\
 medium &      &  $Y < 1$ & $Y > 1$ &   &  $Y < 1$ & $Y > 1$ & $Y < 1$ & $Y > 1$  \\
 \hline \hline
 \multicolumn{9}{c}{PAIRS} \\
 \hline
  homogeneous &
  $\Phi^{8/3} t^{-1/3}$ & $\Phi^2 t^{-1/3}$      & $\Phi^{7/3} t^{-1/3}$ & $\Phi t^{-3/4}$ & 
    $\Phi^2 t^{-3/4}$   & $\Phi^{3/2} t^{-3/4}$  & $\Phi t^{-1}$         & $\Phi^{1/2} t^{-1}$ \\
  $n \propto r^{-2}$ wind &
  $\Phi^{8/3} t^{-5/3}$ & $\Phi^2 t^{-1}$        & $\Phi^{7/3} t^{-5/3}$ & $\Phi t^{-5/4}$ & 
    $\Phi^2 t^{-9/4}$   & $\Phi^{3/2} t^{-5/4}$  & $\Phi t^{-2}$         & $\Phi^{1/2} t^{-1}$ \\
  \hline   \hline
 \multicolumn{9}{c}{FORWARD SHOCK} \\
 \hline
  homogeneous &
  $t^{1/2}$   & \multicolumn{2}{c}{$t^{1/6}$}    & $t^{-3/4}$ & \multicolumn{2}{c}{$t^{-1/4}$}  & 
        \multicolumn{2}{c}{$t^{-1}$} \\
  $n \propto r^{-2}$ wind &
  $t^0$       & \multicolumn{2}{c}{$t^{-2/3}$}   & $t^{-5/4}$ & \multicolumn{2}{c}{$t^{-1/4}$}  & 
        \multicolumn{2}{c}{$t^{-1}$} \\
 \hline \hline
 \multicolumn{9}{c}{REVERSE SHOCK (adiabatic cooling)} \\
 \hline
  homogeneous &
  $t^{-0.31}$ & \multicolumn{2}{c}{$t^{-0.31}$}  & $t^{-1.11}$ & \multicolumn{2}{c}{$t^{-2.5}$} & 
        \multicolumn{2}{c}{$t^{-2.5}$} \\
  $n \propto r^{-2}$ wind &
  $t^{-0.28}$ & \multicolumn{2}{c}{$t^{-0.28}$}  & $t^{-1.46}$ & \multicolumn{2}{c}{$t^{-2.5}$}  & 
        \multicolumn{2}{c}{$t^{-2.5}$} \\
 \hline \hline
\end{tabular}
\end{table*}

 Table 1 illustrates the un-surprising correlation of the pair flux with the fluence $\Phi$ of 
the high-energy photons that form the pairs. For the same distribution of leptons with energy, 
the pair synchrotron emission always decays faster than the forward-shock's (taking into account 
that $\Phi \propto t^{-1.3\pm 0.3}$). Compared with the synchrotron emission from the reverse-shock,
after that shock crossed the ejecta shell and electrons cool adiabatically, the pair emission
decay is steeper at $\nu < \nu_p$ and slower at $\nu > \nu_b$. 

 2) $\Gamma < \Gamma_\tau$ and $nb > (nb)_\tau$. 
In this case, $\tpm > 1$, $\gamma_c = 1$, the scattering optical thickness of hot pairs is $\tau_h < 1$,
i.e. most scatterings are on cold pairs and leave the synchrotron photon energy unchanged.
Scatterings by cold leptons increase the optical thickness to self-absorption to an effective 
value $\ttau_a = \sqrt{\tau_a(\tau_a+\tpm)}$. The observer receives emission from a layer of 
geometrical thickness $l'_a$ corresponding to one self-absorption optical thickness, 
$l'_a = \Delta'/\ttau_a$. Because we take into account only photons that escape the pair medium
in one dynamical timescale (on which the number of pairs changes), i.e. only photons from a 
scattering optical depth $\sqrt{\tpm}$, the received synchrotron flux is
\begin{equation}
 F_\nu = \frac{F_\nu^{(o)}}{ \max [\sqrt{\tpm}, \ttau_a (\nu) ] } 
   \quad \ttau_a(\nu) \equiv  \sqrt{\tau_a(\nu) [\tau_a(\nu)+\tpm]}
\label{Fnu2}
\end{equation}
or
\begin{equation}
 F_\nu = F_\nu^{(o)} \left\{ \begin{array}{ll} 
    \tpm^{-1/2} & (\tau_a < 1) \\ (\tau_a \tpm)^{-1/2} & (1 < \tau_a < \tpm) \\ 
    \tau_a^{-1} & (\tpm < \tau_a) \\ \end{array} \right.
\end{equation}

3) $\Gamma < \Gamma_\tau$ and $(nb)_\tau > nb > (nb)_c$.
In this case, $\tpm > 1$, $\gamma_c =1$, and $\tau_h > 1$. Because the upscattering of a synchrotron
photon by hot pairs means the "destruction" of the synchrotron photon (which will be counted as an
inverse-Compton photon), upscatterings reduce the emergent synchrotron flux in a fashion similar to
self-absorption
\begin{equation}
 F_\nu = \frac{F_\nu^{(o)}}{ \max[ \sqrt{\tpm}, \ttau_a(\nu), \ttau_h ] }  \quad (\tpm > 1)
\label{Fnu3}
\end{equation}
\begin{equation}
  \ttau_h \equiv \sqrt{\tau_h \tpm} \;,\; 
   \ttau_a(\nu) \equiv  \sqrt{\tau_a(\nu) [\tau_a(\nu)+\tpm_c]}
\end{equation}
where $\tau_c = \tau -\tau_h$ is the scattering optical thickness of cold pairs.
Equation (\ref{Fnu3}) reduces to (\ref{Fnu2}) for $\tau_h < 1$.

 The self-absorption frequency $\tnu_a$ defined by $\ttau_a (\tnu_a)= 1$ satisfies 
$[\tpm_c + \tau_a(\tnu_a)] \tau_a(\tnu_a) = 1$, which implies that $\tau_a (\tnu_a) < 1$.
Adding that $\tau_a (\nu)$ decreases with $\nu$, this means that $\tnu_a > \nu_a$, where
$\nu_a$ is the self-absorption frequency without scatterings, defined by $\tau_a (\nu_a) = 1$.
Obviously, scatterings by cold leptons (without changing the energy of the synchrotron 
photon) increase the self-absorption frequency. For $\tpm_c < \tau_a (\tnu_a)$,
we have $\tau_a (\tnu_a) \simeq 1$, thus $\tnu_a \simeq \nu_a$ (trivial). 
For $\tpm_c > \tau_a (\tnu_a)$, we get $\tpm_c \tau_a(\tnu_a) \simeq 1$, which can be solved 
easily: the solution is that of equation (\ref{ga}) but with $\tpm_c \tau_p$ instead of $\tau_p$.
The boundaries involving $\nu_a$ shown in Figure 1 remain unchanged for $\tpm > 1$ because 
$\gamma_c > 1$, hence $\tpm_c = 0$ (there are not any cold pairs).

4) $\Gamma < \Gamma_\tau$ and $nb < (nb)_c$.
In this case, $\gamma_c > 1$, $\tau_c = 0$, and $\tau_h = \tpm > 1$, thus
\begin{equation}
 F_\nu = \frac{F_\nu^{(o)}}{\max [\tau_a(\nu),\tpm] }  
\label{Fnu4}
\end{equation}
Note that equation (\ref{Fnu3}) reduces to (\ref{Fnu4}) for $\tau_c = 0$, thus equation (\ref{Fnu3})
applies to all $\tau > 1$ cases.

 For the sake of generality, equations (\ref{Fnu1}) and (\ref{Fnu3}) can be combined as
\begin{equation}
 F_\nu = \frac{F_\nu^{(o)}}{ \max[ 1, \sqrt{\tpm}, \ttau_a(\nu), \ttau_h, \tpm/3 ] } 
\label{Fnu5}
\end{equation}
valid for any $\tpm$ and $\tau_h$. The last term in the denominator accounts for the formation 
of pairs ahead of the diffusing photon. As discussed before, a photon starting from a scattering 
optical depth $\tau_\o$ will undergo $\tau_\o^2$ scatterings before it diffuses over a distance 
$L'_d = \tau_\o l'_s$ (in the comoving-frame) and exits the pair shell. These scatterings take a 
time $\delta t' = \tau_\o^2 l'_s/c$, hence the photon diffuses at average speed $v'_d = L'_d/\delta t' 
= c/ \tau_\o$ relative to the shocked fluid. For the photon to really escape the pair medium, 
that diffusion speed should exceed the velocity $v'_{rel}$ of the outer edge of the pair-medium
relative to the shocked fluid.
When the pair medium is optically thick, most pairs form within the shocked fluid that produced
the pair-forming photons, whose outer edge is the forward-shock, which moves at speed $v'_{sh} = c/3$ 
relative to the post-shock fluid. Thus, the condition $v'_d > v'_{sh}$ implies that only photons
from optical depth $\tau_\o < 3$ escape the pair medium.

\vspace*{2mm}
\subsection{Inverse-Compton} 

\begin{figure*}
\centerline{\psfig{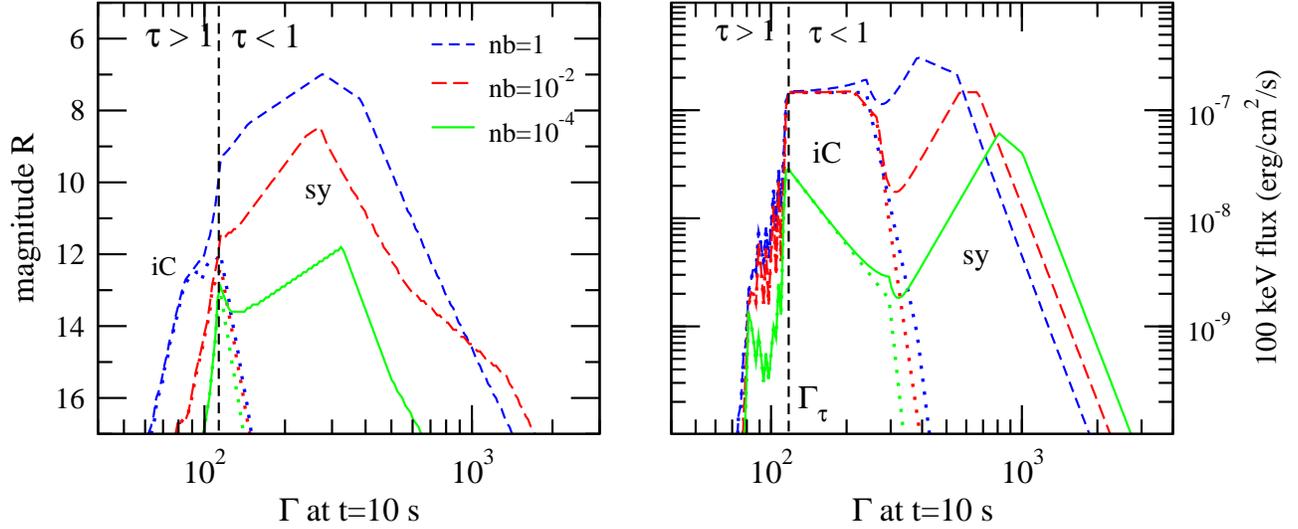}}
\figcaption{ Optical (left panel) and hard X-ray (right panel) emission from pairs formed
    behind the forward-shock from high-energy photons of fluence $\Phi = 10^{-5} \ergcm2$
    above 100 MeV, with a $C_\veps \propto \veps^{-2}$ photon spectrum, at observer time $t_\o=10$ s
    (during the burst), for various magnetic field parameters $nb$ (legend), and for a range of 
    source Lorentz factors $\Gamma (t_\o)$. 
    Dotted lines identify the inverse-Compton emission (iC).
    For an optically-thin ($\tau < 1$) pair-medium, the optical emission is only synchrotron; 
    for $\tau > 1$, it is mostly iC. In contrast, the 100 keV emission can be iC for $\Gamma >
    \Gamma_\tpm$. The wiggly X-ray inverse-Compton flux for $\tau > 1$ is due to variations 
    of the order of inverse-Compton components that escape the optically-thick medium in one 
    dynamical timescale. 
    A low $\Gamma$ entails a higher $\tau$ and a low flux from the pairs in the layer of 
    optical depth $\sqrt{\tau}$, whose emission escapes within one dynamical timescale. 
    A high $\Gamma$ leads to a high threshold for pair-formation, a low number of pairs formed, 
    hence a dimmer pair emission.}
\end{figure*}

\vspace*{2mm}
\subsubsection{$\tau < 1 \; (\Gamma > \Gamma_\tau)$}

 Only one upscattering occurs during one dynamical timescale, hence $Y=Y_1=(4/3)\gamma_i\gamma_c\tau_h$. 
The upscattering of a fraction $\tpm$ of synchrotron photons implies that the peak of the first 
inverse-Compton spectrum is
\begin{equation}
  F_p^{(1)} = \tau_h F_p^{(sy)}
\end{equation}
with $F_p^{(sy)}$ the flux at the peak of the {\sl emergent} synchrotron spectrum:
\begin{equation}
 F_p^{(sy)} = F_p \left\{ \begin{array}{ll}  
  1 & (\nu_a <\nu_p) \\
 \displaystyle{ \left(\frac{\nu_p}{\nu_a}\right)^{1/2} } & (\nu_p<\nu_a<\nu_b) \\
 \displaystyle{ \left(\frac{\nu_p}{\nu_b}\right)^{1/2} \frac{\nu_b}{\nu_a} } & (\nu_b<\nu_a) \\
   \end{array} \right.
\label{Fpsy}
\end{equation}
with $F_p$ the peak flux of the intrinsic synchrotron spectrum, given in equation (\ref{Fp}), 
$\nu_a$ being the self-absorption frequency, $\nu_p = \min(\nu_i,\nu_c)$ is the peak energy of the 
$F_\nu$ synchrotron spectrum, and $\nu_b = \max(\nu_i,\nu_c)$ is the peak of the $\nu F_\nu$ synchrotron
spectrum. The emergent synchrotron spectrum peaks at $\max (\nu_p, \nu_a)$.

 The spectrum of the first inverse-Compton emission has breaks at  
\begin{equation}
 \nu_a^{(1)} \equiv \frac{4}{3} \gamma_p^2 \nu_a ,\;
 \nu_p^{(1)} \equiv \frac{4}{3} \gamma_p^2 \max (\nu_p,\nu_a) ,\;
 \nu_b^{(1)} \equiv \frac{4}{3} \gamma_b^2 \max (\nu_b,\nu_a) 
\end{equation}
The shape of the {\sl intrinsic} inverse-Compton spectrum is the same as for the synchrotron 
spectrum (equation \ref{Fnu}), but upscattering of the self-absorbed synchrotron emission
$F_\nu \propto \nu^{1/3}/\tau_a(\nu) \propto \nu^2$ yields a softer {\sl emergent} inverse-Compton 
spectrum $F_\nu^{(1)} \propto \nu$, that spectrum being
\begin{equation}
 F_\nu^{(1)} = F_p^{(1)} \left\{ \begin{array}{ll}  
 \left( \nu/\nu_a^{(1)} \right) \left( \nu_a^{(1)}/\nu_p^{(1)} \right)^{1/3} 
              & (\nu<\nu_a^{(1)}<\nu_p^{(1)}) \\
 \nu/\nu_a^{(1)}        & (\nu<\nu_a^{(1)} , \nu_p^{(1)} < \nu_a^{(1)}) \\
 \left( \nu/\nu_p^{(1)} \right)^{1/3}  & (\nu_a^{(1)}<\nu<\nu_p^{(1)}) \\
 \left( \nu_p^{(1)}/\nu \right)^{1/2} & (\nu_p^{(1)} < \nu < \nu_b^{(1)}) \\
 \sqrt{\nu_p^{(1)}\nu_b^{(1)}}/\nu  & (\nu_a^{(1)} < \nu_b^{(1)} < \nu) \\
 \nu_a^{(1)}/\nu   & (\nu_b^{(1)} < \nu_a^{(1)} < \nu) \\     \end{array} \right.
\label{Fic}
\end{equation}

\vspace*{2mm}
\subsubsection{$\tau > 1 \; (\Gamma < \Gamma_\tau)$}

 If the effective optical thickness to scattering by hot pairs $\ttau_h = \sqrt{\tau_h \tpm} < 1$
then the above equations for $\tpm < 1$ apply, but with the self-absorption frequency $\tnu_a$ 
accounting for scatterings by cold pairs instead of $\nu_a$.

 For $\ttau_h >1$, the cooling $\gamma_c$ and Compton $Y$ are those calculated in \S3.2 for $\tau_h$
upscatterings that a synchrotron photon, starting from optical depth $\sqrt{\tpm}$, undergoes during 
one dynamical timescale. 
Taking into account than only photons from a scattering optical depth $\tau_\o < 3$ catch up with 
the forward-shock, and that such photons undergo up to 9 scatterings before escaping the pair medium, 
out of which $9 \tau_h/\tpm$ are upscatterings, it follows that the maximal inverse-Compton order 
to be considered is $iC_{max} = \min(\tau_h,9\tau_h/\tpm)$.
 The emergent upscattered emission is the sum of $iC_{max}$ inverse-Compton components 
peaking at progressively higher energies. After $i$ upscatterings, the scattered photon has diffused, 
on average, a distance $\sqrt{i} l'_s$, where $l'_s = \Delta'/\ttau_h$ is the photon free mean-path
between upscatterings. Therefore, most of the $i$-th inverse-Compton emission arises from a 
upscatterings of seed photons produced by a layer located at upscattering optical depth 
$\sqrt{i-1}\div \sqrt{i}$, which we will call the "$i$-th iC layer". With increasing distance from it, 
inner and outer layers yield a lesser and lesser contribution to the $i$-th inverse-Compton emission.

\begin{figure*}
\centerline{\psfig{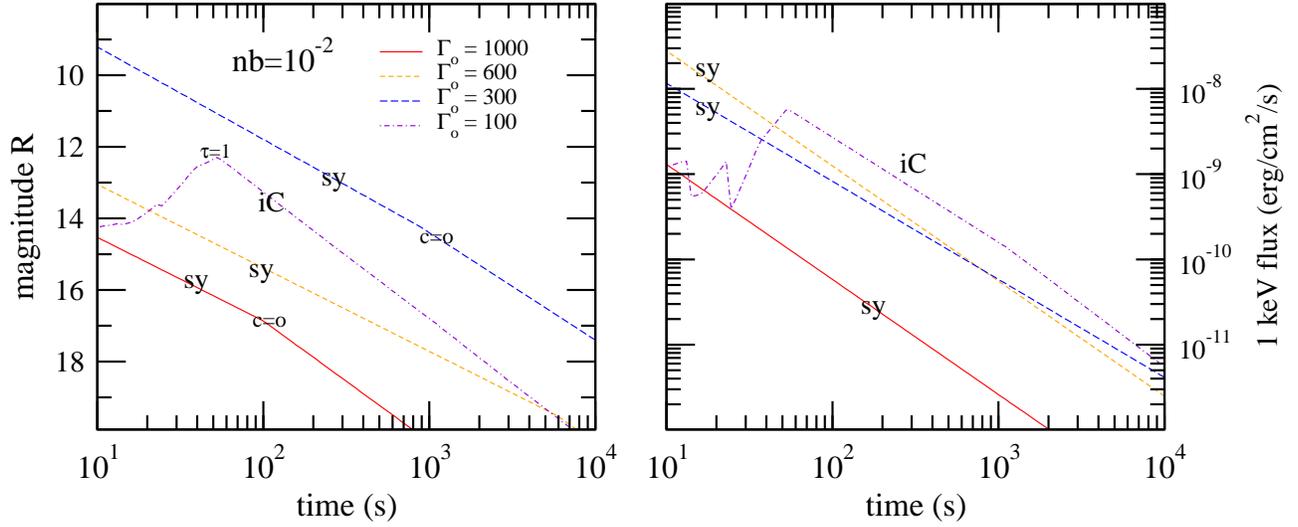}}
\figcaption{ Optical (left) and soft X-ray (right) light-curves from pairs produced by high-energy
      photons with fluence $\Phi = 10^{-5} (t/t_\o)^{-1/3} \ergcm2$ and $C_\veps \propto \veps^{-2}$ 
      photon spectrum, for a fixed magnetic field parameter $nb = 10^{-2}$, and a shock deceleration 
      $\Gamma = \Gamma_\o (t/t_\o)^{-3/8}$ corresponding to a {\sl homogeneous} medium (of particle
      density $n$), starting from various $\Gamma_\o$ at $t_\o = 10$ s. 
      The dominant emission process is indicated, as well as the origin of light-curve breaks
      ("c=o" means cooling frequency crossing the optical). Only for the lowest $\Gamma_\o = 100$,
      the pair medium is optically thick to photon scattering, until the time of the inverse-Compton
      light-curve. }
\end{figure*}

 For ease of calculations and accounting of the seed synchrotron photons, we assume that the $i$-th 
inverse-Compton emission arises from upscatterings of synchrotron photons produced {\sl only} by the 
$i$-th iC layer. Given that all seed photons from $i>1$ layers are upscattered, this one-to-one 
correspondence between inverse-Compton order and optical depth should entail that {\sl all} the 
synchrotron photons produced by the $i$-th iC layer become $i$-th inverse-Compton photons,
thus the peak flux of the $i$-th inverse-Compton spectrum is equal to the peak flux of the 
synchrotron spectrum for the $i$-th iC layer:
\begin{equation}
 F_p^{(i)} = \frac{\sqrt{i}-\sqrt{i-1}}{\ttau_h} F_p^{(sy)} \quad  
  [i=1 \div \min(\tau_h,9\tau_h/\tpm)]
\end{equation}
where $F_p^{(sy)}$ is the flux at the peak of the synchrotron spectrum for the entire pair medium, 
given in equation (\ref{Fpsy}), but with $\tnu_a$ corresponding to self-absorption only in the 
$i$-th iC layer:
\begin{equation}
 \ttau_a^{(i)} (\tnu_a ) = 1 \;,\; 
  \ttau_a^{(i)} \equiv \sqrt{\tau_a^{(i)} \left(\tpm_c^{(i)} + \tau_a^{(i)} \right)} 
\end{equation}
where $\tpm_c^{(i)} \equiv (\sqrt{i}-\sqrt{i-1}) \tpm_c/\ttau_h$ is the optical thickness 
to scattering by cold pairs in the $i$-th iC layer (although the $i$-th iC layer is 
optically thin to upscatterings by hot pairs, it is not necessarily thin to scatterings 
by cold pairs), and with $\tau_a^{(i)}(\nu)$ as given in equation (\ref{taua1}) but 
for the scattering optical thickness $\tpm^{(i)} \equiv (\sqrt{i}-\sqrt{i-1}) \tpm/\ttau_h$
of the $i$-th iC layer.

 The flux of the $i$-th inverse-Compton emission is as given in equation (\ref{Fic}) but with 
$F_p^{(i)}$ instead of $F_p^{(1)}$ and with spectral breaks at
\begin{equation}
 \nu_a^{(i)} \equiv \left( \frac{4}{3} \gamma_p^2 \right)^i \tnu_a 
\end{equation}
\begin{equation}
 \nu_p^{(i)} \equiv \left( \frac{4}{3} \gamma_p^2 \right)^i \max (\nu_p,\tnu_a) 
\end{equation}
\begin{equation}
 \nu_b^{(i)} \equiv \left( \frac{4}{3} \gamma_b^2 \right)^i \max (\nu_b,\tnu_a) 
\end{equation}

 To summarize the accounting of photons for $\tpm > 1$: the outermost layer of one upscattering
optical depth ($i=1$) yields the synchrotron emission and the first inverse-Compton scattering,  
layers of geometrical depth $(\sqrt{i-1} \div \sqrt{i})(\Delta/\ttau_h)$ produce all the seed 
photons for the $i$-th inverse-Compton emission, but we ignore the inverse-Compton emission of 
order higher than $\tau_h$, produced by pairs at geometrical depth larger than 
$\sqrt{\tau_h} \Delta/\ttau_h = \Delta/\sqrt{\tpm}$ because that emission is trapped in the 
pair medium for longer than one dynamical timescale, on which change the number of pairs and their
distribution with energy.

\vspace*{2mm}
\section{Optical and X-ray light-curves} 
\label{results}

\begin{figure*}
\centerline{\psfig{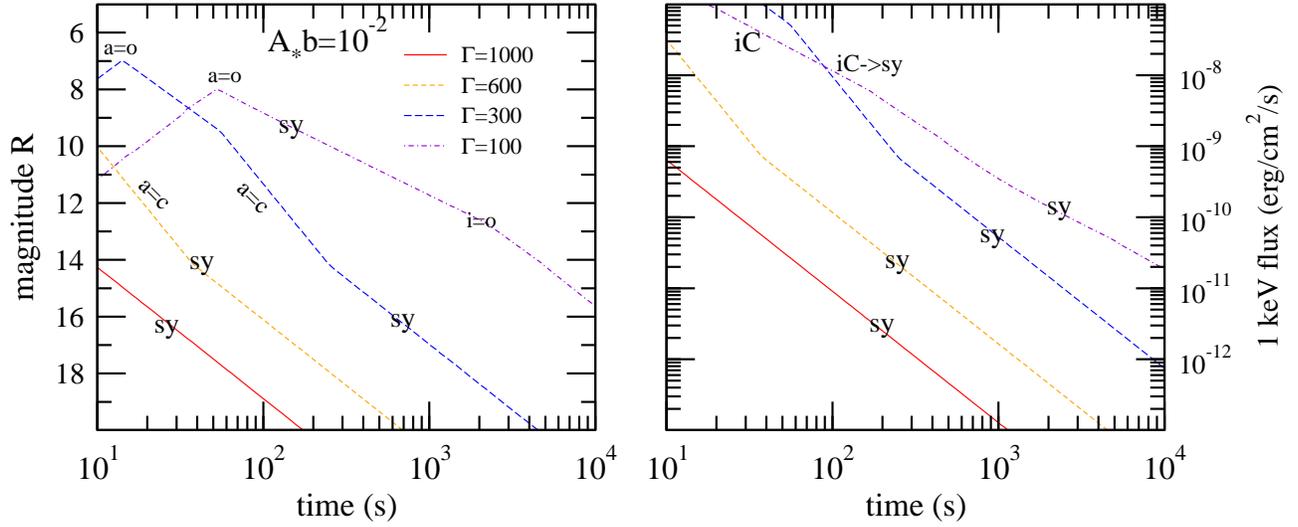}}
\figcaption{ As for Figure 3 but for a source deceleration $\Gamma = \Gamma_\o (t/t_\o)^{-1/4}$ 
      corresponding to a {\sl wind-like} medium of proton density $n(R) = 3.10^{35}\, A_*/R^2$,
      and a magnetic field parameter $A_* b = 10^{-2}$.
      Light-curve breaks originate from a spectral break crossing the optical ("a=o" for absorption, 
     "i=o" for injection), from the occurrence of the ($\gamma_a = \gamma_c$, $Y < 1$) regime
      ("a=c"), or from an inverse-Compton to synchrotron transition ("iC$\rightarrow$sy", 
      in the X-ray). }
\end{figure*}

 The formalism presented so far allows the calculation of the optical and X-ray synchrotron 
self-Compton flux from pairs at the observer-time $t$ when LAT measured a fluence $\Phi (t)$, 
for an assumed source Lorentz factor $\Gamma$, and a shock magnetic field $B$ (parametrized
and tied to $\Gamma$ by the $nb$ parameter).
 
 The "pseudo light-curves" of Figure 2 show the synchrotron and inverse-Compton flux at 2 eV
(optical) and 100 keV (hard X-ray) as function of the source Lorentz factor $\Gamma$ at $t = 10$ s
(during the prompt emission phase), for a LAT fluence $\Phi = 10^{-5} \ergcm2$, and a
few values for the magnetic field parameter $nb$. 
 The brightest optical flux is obtained when the peak frequency $\nu_p$ of the synchrotron 
spectrum is close to the optical range. We note that the peak flux of equation (\ref{Fp}) 
means magnitude $V=6.6$, comparable to that shown in Figure 2 for $nb =1$ and with the 
brightest optical counterpart (emission during the prompt/burst phase) ever observed
(GRB 080319B, reached peak magnitude $V = 5.3$ at 20 s after trigger -- Greco et al 2009). 
 
 The 10 keV--1 MeV fluence of Fermi-GBM bursts is about ten times larger than the 100 MeV--10 GeV 
fluence of the corresponding LAT afterglows, thus the average MeV flux of a 10 s burst is
$F_{grb} \sim 10^{-5} {\;\rm erg\, cm^{-2}\, s^{-1}}$. 
Figure 2 shows that, during the prompt phase, the pair emission 
at 100 keV is dimmer by about two orders of magnitude than the burst.

 Figure 3 shows proper pair-emission optical and soft X-ray light-curves for a source deceleration 
corresponding to a blast-wave interacting with a homogeneous ambient medium (uniform $n$) --
$\Gamma \propto t^{-3/8}$ -- a slowly-decreasing high-energy fluence $\Phi$, a constant magnetic field 
parameter $nb$, and starting from various initial Lorentz factors $\Gamma_\o$ at $t_\o = 10$ s. 
The shock deceleration corresponds to moving from right to left, on a horizontal line in Figure 1. 
The pair flux is calculated at each epoch in same way as for Figure 2, adjusting the fluence $\Phi$. 
For Figure 4, the ambient medium has a wind-like structure, $n \propto R^{-2}$, as for a massive star 
GRB progenitor, hence the shock deceleration is $\Gamma \propto t^{-1/4}$ and the magnetic field 
parameter evolves as $nb \propto t^{-1}$, the deceleration corresponding
to moving from right to left on a $nb \propto \Gamma^4$ line in Figure 1.

 Because the pre-deceleration phase of the forward-shock and the rise of the high-energy flux $\Phi$ 
before $t_\o$ are not accounted for, the calculated pair synchrotron light-curves miss the initial 
rising part, but they can still display peaks when the self-absorption $\nu_a$ break or the $F_\nu$
peak $\nu_p = \min(\nu_i,\nu_c)$ fall below the observing band. Otherwise, crossing by the 
$\nu F_\nu$ peak frequency $\nu_b = \max(\nu_i,\nu_c)$ yields a light-curve break. 

 For a typical LAT light-curve, $\Phi(t) = Ft \propto t^{-1/3}$, the pair flux has a power-law decay 
$F_\nu \propto t^{-\alpha}$ with $\alpha \in (1.0,1.4)$ for a homogeneous medium and $\alpha \in 
(1.2,2.0)$ for a wind, the latter range being more compatible with the observations of GRB optical
counterparts (flashes): $\alpha =1.8$ for GRB 990123 (Akerlof et al 1999), 
$\alpha =2.0$ for GRB 061126 (Perley et al 2008), 
$\alpha =2.5$ for GRB 080319B (Wozniak et al 2009), 
$\alpha =1.7$ for GRB 130427A (Vestrand et al 2014). 
Thus, a wind-like medium is favored when interpreting the optical flashes as emission from pairs;
still, we note that steeper decays of the pair emission result for a fluence $\Phi$ decreasing faster 
than $t^{-1/3}$.

 While Figures 3 and 4 show that the X-ray flux from pairs at 0.1--10 ks can be compatible with
that measured for Swift X-ray afterglows -- $F_\nu = 10^{-11}-10^{-10} {\;\rm erg\, cm^{-2}\, s^{-1}}$ 
(O'Brien et al 2006) -- the above pair light-curve decays are too steep compared with X-ray plateau
measurements -- $\alpha \in (0.2,0.8)$ -- thus, only the faster-decaying plateaus can be accommodated
by pair emission, provided that the fluence $\Phi$ is nearly constant, and that the ambient medium is
homogeneous.

\vspace*{2mm}
\subsection{Brightest optical flash from pairs} 

 Figure 5 shows the maximal optical flux from pairs at $t = 10$ s (i.e. during the prompt emission phase),
for a range of high-energy fluence, obtained by searching a reasonable range of the $(\Gamma,nb)$
parameter space: $\Gamma \in (10,10^4)$ and $nb \in (10^{-5},10^5)$. As illustrated in Figure 2,
one expects that $\Gamma \in (100,1000)$ for the brightest optical flash, because too low Lorentz factors
lead to optically-thick pairs and less radiation escapes in a dynamical timescale, while too high
Lorentz factors increase the pair-formation threshold energy, leading to the formation of fewer pairs,
and reducing the flux produced by pairs. As suggested by Figure 2, Figure 5 shows that the maximal
optical synchrotron emission from pairs is larger than that from inverse-Compton, at any fluence.

 The important result shown in Figure 5 is the existence of an upper limit $R_{\min}$ on the brightness 
of the optical flash from pairs, which offers test of this model: any flash brighter than shown in Figure 5,
for the measured LAT fluence, cannot originate from pairs. In more detail, the brightest observed
LAT afterglows ($\Phi \sim 10^{-4} \ergcm2$) could yield optical flashes as bright as $R = 5$, the dimmest 
LAT afterglows ($\Phi \sim 10^{-6} \ergcm2$) may produce $R = 9$ optical flashes, with reasonably bright 
flashes originating from the pairs produced in GRB afterglows that are not detectable by LAT above 100 MeV.

 We note that the existence of $R_{\min}$ is not a immediate consequence of equation (\ref{Fp}), 
which leaves the possibility of a brighter optical flash than shown in Figure 5, if the peak frequency 
$\nu_p$ of the synchrotron spectrum fell in the optical and if it were not self-absorbed.
Instead, the brightest optical flash from pairs shown in Figure 5 corresponds to $\nu_p$ being
below optical ($\nu_o$); more exactly, $\nu_c < \nu_a = \nu_o < \nu_i$ (and $Y > 1$, $\tau < 1$)
is satisfied everywhere on the $R_{\min} (\Phi)$ line, a condition also fulfilled by the brightest
two "peaks" shown in Figure 2 (left panel). 

\begin{figure*}
\centerline{\psfig{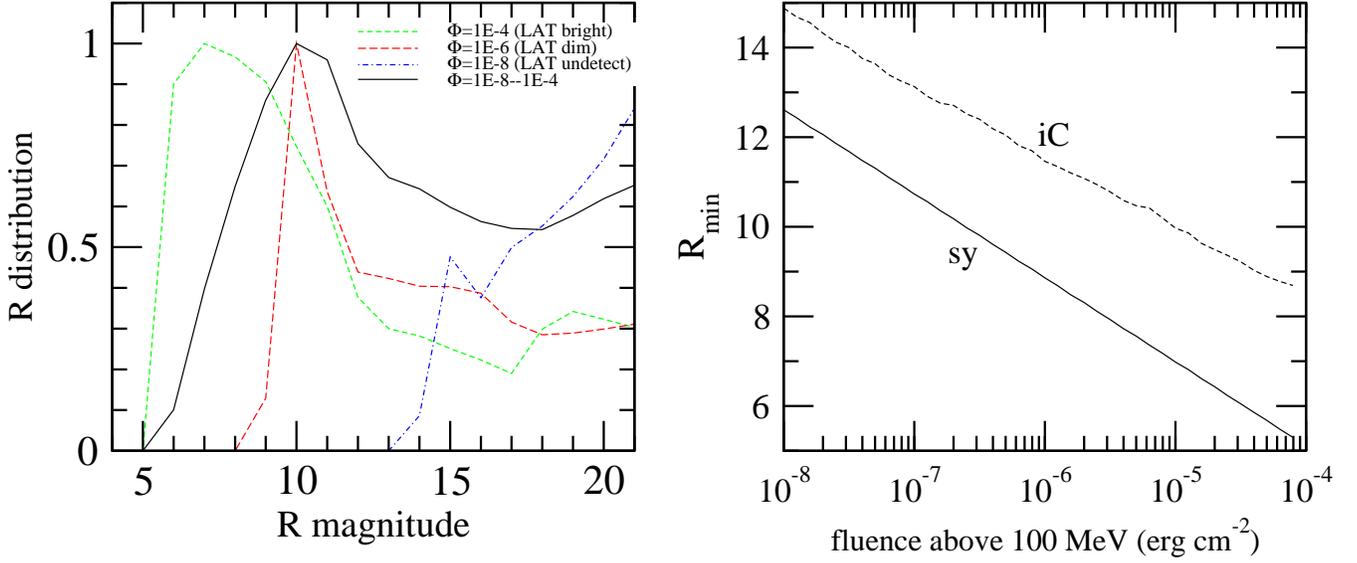}}
\figcaption{ {\sl Left panel}: distribution (number of events in $\delta R = 1$ bin, relative to maximum)
     of pair optical flash magnitude $R$ at $t = 10$ s, for three fixed GeV fluences $\Phi$ (corresponding 
     to a bright, dim, and undetectable LAT emission) and for a
     GeV fluence distributed uniformly in log-space (solid line). The source magnetic field parameter $nb$
     and Lorentz factor $\Gamma$ at time $t$ were assumed uniformly distributed in log-space, spanning 
     the range $10^{-4}-10^4\, {\rm cm^{-3}}$ and $10^2-10^4$, respectively (the shape of the distributions 
     and its peak location depend on the assumed range and distribution of these two parameters, but the 
     brightest end magnitude depends only on the fluence).
   {\sl Right panel}: brightest pair optical flash flux $R_{\min}$ (i.e. the bright end of the distributions
     shown in the left panel) as a function of fluence $\Phi$, with the two parameters $(\Gamma,nb)$ left 
     free.  The $R_{\min}$ is reached for $\nu_c < \nu_a < \nu_i$, $Y > 1$, $\tau < 1$, and when the 
     self-absorption frequency is in the optical. 
     For this ordering of synchrotron break frequencies, $R_{\min}$ is independent of the burst redshift 
     and very weakly dependent on the epoch $t$, thus the upper limit $R_{\min}$ on the optical flash 
     flux shown here is a robust prediction for the synchrotron emission from pairs. } 
\end{figure*}

 The surprising aspect of the brightest optical flash shown in Figure 5 is that $R_{\min} (\Phi)$ is
{\sl independent} of the burst redshift $z$. Perhaps a moderate dependence of $R_{\min}$ on redshift
is expected because $R_{\min}$ is calculated for a fixed fluence $\Phi$, thus a higher $z$ implies
a larger afterglow output above 100 MeV, a larger number of formed pairs, and a larger synchrotron
luminosity that compensates for the larger luminosity distance. That $R_{\min} (\Phi)$ is $z$-independent
can be proven in the following way. From equations (\ref{Fp}) -- (\ref{Fnu}), it follows that, 
for $\nu_c < \nu_a < \nu_o < \nu_i$, the optical flux satisfies above self-absorption is
\begin{equation}
 F_o = F_p \left( \frac{\nu_c}{\nu_o} \right)^{1/2} \propto 
       \z^{5/2} \frac{\Phi^{3/2} (nb)^{1/4}}{\Gamma^2 t^{3/2}} \propto \Gamma^{-2}
\label{Fo1}
\end{equation}
For $\nu_c < \nu_a < \nu_i$, equations (\ref{nui}) and (\ref{taua})--  (\ref{ga}) lead to
\begin{equation}
 \gamma_a = \gamma_c \tau_p^{1/6} \propto \left( \frac{\tau \gamma_c}{B} \right)^{1/6} ,
  \quad \nu_a \propto \z^{5/3} \frac{\Phi^{1/2} (nb)^{1/6}}{\Gamma^{4/3} t^{7/6}} 
\label{nua}
\end{equation}
From equations (\ref{taua1}) and (\ref{Fp}) -- (\ref{Fnu1}), the optical flux below $\nu_a$ is
\begin{equation}
 F_o = F_p \left( \frac{\nu_c}{\nu_a} \right)^{1/2}  \left( \frac{\nu_o}{\nu_a} \right)^{5/2} 
      \propto \z^{-5/2} \frac {\Gamma^2 t^2} {\Phi^0 (nb)^{1/4}} 
\label{Fo2}
\end{equation}
Defining $\Gamma_p$ by $\nu_a (\Gamma_p) = \nu_o$, it follows from equation (\ref{nua}) that
$i)$ $\nu_o < \nu_a$ for $\Gamma < \Gamma_p$, hence $F_o \propto \Gamma^2$ (equation \ref{Fo2}),
the optical flux increasing with $\Gamma$, and 
$ii)$ $\nu_a < \nu_o$ for $\Gamma > \Gamma_p$, hence $F_o \propto \Gamma^{-2}$ (equation \ref{Fo1}),
the optical flux decreasing with $\Gamma$.
Consequently, the optical flux is maximal for $\Gamma = \Gamma_p$, with $\Gamma_p \propto 
\z^{5/4} \Phi^{3/8} (nb)^{1/8} t^{-7/8}$ following from the defining condition $\nu_a (\Gamma_p) = 
\nu_o$. Substituting $\Gamma_p$ in either equation (\ref{Fo1}) or (\ref{Fo2}), the maximal
optical flux satisfies
\begin{equation}
  F_o^{\max} \equiv F_o (\Gamma_p) \propto \z^0 (nb)^0 \Phi^{3/4} t^{1/4}
\label{Fo}
\end{equation}
being independent of redshift and also of the magnetic field parameter $nb$ (meaning that, 
for any $nb$ that allows $\nu_c < \nu_a = \nu_o < \nu_i$, $Y > 1$, and $\tau < 1$, there is 
a $\Gamma_p$ that maximizes the optical flux to the same value $F_o (\Phi,t)$). 
The coefficient missing from equation (\ref{Fo}) can be determined by carrying the 
coefficients of all equations involved in its derivation. From Figure 5, the maximal optical 
flux of equation (\ref{Fo}) is
\begin{equation}
  R_{\min} = 8.7 - 2.5 \log \left( \frac{F_o^{\max}}{\rm Jy} \right) = 
    7.0 - \frac{15}{8} \log \Phi_{-5} - \frac{5}{8} \log t_1
\label{Rmin}
\end{equation}

 For GRB 130427A, the optical flash peaked at $R \simeq 7.4$ at 10--20 s after trigger (Wren et al 2013),
when the LAT 0.1--100 GeV fluence was $\Phi \simeq 4.10^{-5} \ergcm2$ (Tam et al 2013). The upper limit 
given in equation (\ref{Rmin}), $R_{\min} \simeq 5.8$, is brighter than measured, thus a pair origin 
for the optical flash of GRB 130427A is not ruled out.

\vspace*{2mm}
\subsection{Caveats} 

 Our assumption that the power-law spectrum of the high-energy photons (measured by LAT at 
100 MeV--10 GeV) extends well outside that range could lead to an {\sl overestimation} of 
the emission from pairs in two ways. 

 Equations (\ref{nui}) and (\ref{B}) show that the synchrotron emission at photon energy $h\nu$ 
is produced by pairs of shock-frame energy $\gamma (\nu) = 155\, (\z h\nu/1 eV)^{1/2} (nb)^{-1/4} 
\Gamma_2^{-1}$. Pairs of this energy are formed from photons of observer-frame energy $\veps (\nu) 
\simeq 2 \gamma \Gamma m_e c^2/(z+1) = 5.3\, (h\nu/1 eV)^{1/2} \z^{-1/2} (nb)^{-1/4}$ GeV. Thus, 
optical synchrotron emission requires pairs formed from seed photons of $\veps \simg 10$ GeV
(which have been occasionally detected by LAT), 1 keV synchrotron emission requires photons of 
$\veps \simg 200$ GeV (the highest-energy photon detected by LAT had $\sim 100$ GeV, for the 
$z=0.3$ GRB 130427A -- Fan et al 2013, Tam et al 2013), while 100 keV synchrotron emission 
requires photons of energy $\veps \simg 3$ TeV. 
 
 At the other end, if the high-energy spectrum 
has a break not far below 100 MeV, the assumption of a single power-law overestimates the 
number of target photons for the $\veps (\nu)$ photon, which leads to an overestimation of the 
optical thickness to pair-formation and of the number of $\gamma (\nu)$ pairs that are formed 
(if the true optical thickness $\taugg [\veps(\nu)]$ is below unity). 
From equation (\ref{ethr}), the pair-formation threshold-energy for a $\veps (\nu)$ photon is 
$\veps_t (\nu) = 0.22 (h\nu/1 eV)^{-1/2} \z^{-3/2} (nb)^{1/4} \Gamma_2^2$ MeV, which, for both 
optical and X-ray photons, is well below the lower edge of the LAT window.

 Not accounting for the decollimation introduced by the scattering of the seed photons on pairs 
leads to an {\sl underestimation} of true number of pairs. An estimate of the importance of that pair-cascade 
can be obtained by first noting that most pairs are formed by the more numerous, lower energy photons
above the threshold for pair-formation, i.e. by the $\sim 1$ MeV photons of the LAT spectrum extrapolation.
For a flat $\nu F_\nu$ LAT spectrum, the fluence of the 1 MeV photons is comparable to the LAT fluence,
hence the 1 MeV photons should have an energy output of $10^{53 \pm 1}$ erg. For a $10^{53}$ erg burst
lasting for 10 s, the outflow pair-loading through a cascade process was shown (Kumar \& Panaitescu 2004)
to produce a significant pair-enrichment up to a radius $R_{max} = 5\times 10^{15}$ cm. For an
observer-frame time $t$, this radius corresponds to source Lorentz factor $\Gamma_{max} = [(z+1) 
R_{max}/ct]^{1/2} \simeq 220\, (\z/t_1)^{1/2}$. Thus, ignoring the pair-cascade process implies an 
underestimation of the pair number for a source with $\Gamma < \Gamma_{max} = 1.5 (\z^{0.7} \phi5^{0.2} 
t_1^{0.1})^{-1} \Gamma_\tau$, i.e. the pair-cascade is effective mostly when the pair-medium produced 
by the unscattered seed photons is optically-thick ($\tau > 1$). As illustrated in Figure 2, a larger 
number of pairs and the associated larger optical thickness $\tau$ lead to a dimmer emergent emission, 
hence the pair-cascade would reduce even more the pair emission.

\vspace*{2mm}
\section{Conclusions}

 The above investigation of the broadband emission from pairs shows that the synchrotron emission
from pairs formed from $>100$ MeV afterglow photons can accommodate the brightest optical counterparts
(flashes) that were observed (in a few cases) during the prompt (GRB) phase, with the fast decay of 
optical flash pointing to a wind-like circumburst medium or to a faster decaying fluence of the high-energy
photons. The inverse-Compton emission from pairs may yield an afterglow brightening, as seldom seen 
in optical afterglow light-curves. 

 A brighter Fermi-LAT afterglow implies more pairs that can form and, thus, a brighter synchrotron
and inverse-Compton emission from those pairs. The light-curve scalings given in Table 1 quantify 
the positive correlation between the synchrotron pair flux and the high-energy photon fluence $\Phi$.
 In addition to the 100 MeV fluence, the pair emission depends also on the source Lorentz factor
$\Gamma$ and on the magnetic field in the pair medium. A very high Lorentz factor ($\Gamma \simg 1000$) 
raises the threshold energy for pair-formation, reduces the number of pairs and, implicitly, their
emission. A very low Lorentz factor ($\Gamma \siml 50$) leads to a larger number of pairs, an 
optically-thick pair medium, which traps the emission and, consequently, yields a dim pair flux. 
Leaving free the source Lorentz factor $\Gamma$ and the magnetic field parameter $nb$, we find
that the brightest optical flash produced by pairs satisfies equation (\ref{Rmin}), which gives
an upper limit to that optical flash that depends mostly on the high-energy fluence, is weakly 
dependent on the epoch of observation, and, most important, is independent of the source redshift. 
We note that the brightest optical flash from pairs for a given LAT fluence is obtained when the 
self-absorption frequency is in the optical, thus the intrinsic synchrotron optical spectrum of 
the brightest optical flash from pairs should be flat.

 Given that a powerful source of high-energy photons is needed to produce enough pairs that can
account for optical flashes/counterparts and that the source of high energy photons is, most likely, 
the forward-shock (Kumar \& Barniol Duran 2009), the multiwavelength data of GRB afterglows at early 
times should be interpreted/modeled as the sum of synchrotron and inverse-Compton emissions from 
the forward-shock, the reverse-shock, and from pairs, at least in those cases where LAT measures a 
a bright high-energy afterglow. To calculate the pair emission requires three parameters: the magnetic 
field parameter $nb$ (which is also constrained by fitting the multiwavelength afterglow data with
the reverse/forward-shock emission), the LAT afterglow fluence $\Phi$ (which is the blast-wave emission), 
and the source Lorentz factor $\Gamma$ (which is constrained directly by the double-shock emission 
fits at the deceleration time, and indirectly, through the ratio of blast-wave energy to external 
density $E/n$, after deceleration onset). Thus, the pair emission dose not entail any new parameters.
We note that lower limits on $\Gamma$ during the burst phase (pre-deceleration) can be obtained
from the photon-photon opacity of the prompt LAT emission (e.g. Abdo et al 2009).

 The high LAT fluence of GRB 130427A (Ackermann et al 2014), its bright
optical counterpart (Vestrand et al 2014), and broadband coverage would make it a good candidate 
for such a study. The low circumburst medium density, inferred by modeling its 10 s--10 d radio, 
optical, X-ray, and 100 MeV measurements, with synchrotron emission from the reverse and forward
shocks (Laskar et al 2013, Panaitescu et al 2013, Perley et al 2014), implies a very high afterglow 
Lorentz factor ($\Gamma_\o \sim 800$ at 10 s) and, consequently, a low number of formed pairs, 
leading to an optical emission from pairs that is well below the bright optical flash of GRB 130427A.

 The pair emission discussed here offers an alternate explanation (to the reverse-shock) for GRB optical
counterparts. Both "mechanisms" can yield bright optical flashes with a fast decay after the peak.
The pair optical flux should be correlated with the GeV contemporaneous fluence, but that feature
may exist also for reverse-shock flashes, if most of the GeV flux arises from the reverse-shock.
The post-peak decay of the optical pair flux, given in Table 1 for a $\veps^{-2}$ LAT spectrum,
can be generalized and used to test a pair origin for the optical flash, although a continuous 
reverse-shock (i.e. not one that ceases at the optical peak time, followed by adiabatic cooling
of ejecta electrons) may yield a similar decay index -- spectral slope closure relation. Perhaps,
modeling of the broadband afterglow data will be the best way to disentangle the pair emission 
and reverse-shock emissions.

\acknowledgments{This work was supported by an award from the Laboratory Directed Research and 
   Development program at the Los Alamos National Laboratory}


\begin{references}
\reference{} Abdo A. et al, 2009, Science 323, 1688
\reference{} Akerlof C. et al, 1999, Nature 398, 400
\reference{} Ackermann M. et al, 2013, ApJS 209, 11
\reference{} Ackermann M. et al, 2014, Sci 343, 42
\reference{} Fan Y.-Z. et al, 2013, ApJ 776, 95
\reference{} Greco G. et al, 2009, Mem. S. A. It. 80, 231
\reference{} Kumar P., Panaitescu A., 2000, ApJ 541, L51
\reference{} Kumar P., Panaitescu A., 2004, MNRAS 354, 252
\reference{} Kumar P., Barniol Duran R., 2009, MNRAS 409, 226 
\reference{} Laskar T. et al, 2013, ApJ 776, 119 
\reference{} O'Brien P. et al, 2006, ApJ 647, 1213
\reference{} Perley D. et al, 2008, ApJ 672, 449 
\reference{} Perley D. et al, 2014, ApJ 781, 37
\reference{} Panaitescu A., Kumar P., 2000, ApJ 543, 66
\reference{} Panaitescu A., Kumar P., 2004, MNRAS 350, 213
\reference{} Panaitescu A., Vestrand T., Wozniak P., 2013, MNRAS 436, 3106
\reference{} Panaitescu A., Vestrand T., Wozniak P., 2014, ApJ 788, 70
\reference{} Rybicki G., Lightman A., 1979, "Radiative Processes in Astrophysics", John Wiley \& Sons: New York
\reference{} Tam P.-H. et al, 2013, ApJ 772, L4
\reference{} Vestrand T. et al, 2014, Science 343, 38
\reference{} Wozniak P. et al, 2009, ApJ 691, 495
\reference{} Wren J. et al, 2013, GCN \#14476 
\end{references}
\end{document}